\documentclass[journal]{IEEEtran}

\usepackage[english]{babel}

\usepackage[letterpaper,top=2cm,bottom=2cm,left=2cm,right=2cm,marginparwidth=1.75cm]{geometry}

\usepackage{subcaption}
\usepackage{amsmath}
\usepackage{graphicx}
\usepackage[colorlinks=true, allcolors=blue]{hyperref}

\usepackage{epstopdf}

\usepackage{varioref}
\usepackage{float}
\usepackage{comment}

\usepackage{xcolor}
\usepackage[linesnumbered,ruled,vlined]{algorithm2e}
\usepackage{amsmath}
\usepackage{caption}
\usepackage{graphicx}
\usepackage{hyperref}

\SetCommentSty{mycommfont}

\SetKwInput{KwInput}{Input}                
\SetKwInput{KwOutput}{Output}              

\usepackage{flushend}
\usepackage{tabularx,booktabs,textcomp}
\newcolumntype{C}{>{\centering\arraybackslash}X} 
\newcolumntype{L}{>{\raggedright\arraybackslash}X} 
\usepackage[noadjust]{cite}

\usepackage{wasysym}
\usepackage{lipsum} 
\usepackage{setspace}
\usepackage{newtxtext,newtxmath}

\usepackage{tikz}
\usepackage{tikz-3dplot}

\usepackage{enumitem}

\usepackage[ruled,vlined,linesnumbered]{algorithm2e}
\usepackage{xcolor}

\title{Maximizing UAV Fog Deployment Efficiency for Critical Rescue Operations}
\author{Abdenacer Naouri, Huansheng Ning, Nabil Abdelkader Nouri,   Amar Khelloufi, Abdelkarim Ben Sada, Salim Naouri, Attia Qammar and Sahraoui Dhelim
	\thanks{Abdenacer Naouri, Huansheng Ning, Amar Khelloufi, Abdelkarim Ben Sada are with the University of Science and Technology Beijing, Beijing 100083, China}
	\thanks{Sahraoui Dhelim is with the School of Computer Science, University College Dublin, Ireland.} 
    \thanks{Nabil Abdelkader Nouri is with the Department of Mathematics and Computer Science, University of Djelfa, Djelfa, Algeria} 
    
 \thanks{Corresponding author: Sahraoui Dhelim (sahraoui.dhelim@ucd.ie).}
}

\begin{document}
\maketitle

\begin{abstract}

In disaster scenarios and high-stakes rescue operations, integrating Unmanned Aerial Vehicles (UAVs) as fog nodes has become crucial. This integration ensures a smooth connection between affected populations and essential health monitoring devices, supported by the Internet of Things (IoT). Integrating UAVs in such environments is inherently challenging, where the primary objectives involve maximizing network connectivity and coverage while extending the network's lifetime through energy-efficient strategies to serve the maximum number of affected individuals. In this paper, We propose a novel model centred around dynamic UAV-based fog deployment that optimizes the system's adaptability and operational efficacy within the afflicted areas. First, we decomposed the problem into two subproblems. Connectivity and coverage subproblem, and network lifespan optimization subproblem. We shape our UAV fog deployment problem as a uni-objective optimization and introduce a specialized UAV fog deployment algorithm tailored specifically for UAV fog nodes deployed in rescue missions. While the network lifespan optimization subproblem is efficiently solved via a one-dimensional swapping method. Following that, We introduce a novel optimization strategy for UAV fog node placement in dynamic networks during evacuation scenarios, with a primary focus on ensuring robust connectivity and maximal coverage for mobile users, while extending the network's lifespan. Finally, we introduce Adaptive Whale Optimization Algorithm (WOA) for fog node deployment in a dynamic network. Its agility, rapid convergence, and low computational demands make it an ideal fit for high-pressure environments.

\end{abstract}

\begin{IEEEkeywords}
Internet of Things, Fog Computing, UAVs, Fog Node Deployment, Multi-objective Optimization, Whale Optimization Algorithm (WHO), Rescue Operations, Network Coverage, Network Connectivity, Network Lifespan.
\end{IEEEkeywords}

\section{Introduction}
\label{sec.1}

Nowadays, the utilization of unmanned aerial vehicles (UAVs) as airborne wireless communication platforms has garnered substantial attention in recent years, primarily due to their remarkable mobility and adaptable deployment capabilities \cite{han2022survey}. In the area of cellular applications, UAVs have found practical applications as temporary base stations (BSs) for offloading task traffic \cite{cheng2018uav}, facilitating disaster recovery efforts \cite{hydher2020intelligent}, and serving as relays to extend network coverage to remote and underserved regions \cite{zeng2016throughput}. Furthermore, within the domain of the Internet of Things (IoT) and wireless sensor networks (WSN), UAVs have emerged as versatile assets, functioned as mobile aggregators or sink nodes to streamline data collection processes \cite{motlagh2017uav}. 

Efficiently deploying drones is a crucial challenge in designing systems that use unmanned aerial vehicles (UAVs), as discussed in \cite{koyuncu2018deployment}, \cite{lakew20203d}, \cite{zeng2017energy}. Various optimization methods have been suggested, including evolutionary algorithms \cite{park2020optimized}, \cite{liu2022joint}, \cite{tran2021uav}. In \cite{li2019rechargeable}, the authors take a cooperative approach to provide coverage and long-term data services for IoT devices in UAV-supported networks. The complex problem is divided into three smaller subproblems and use an iterative algorithm based on block coordinate descent to solve these subproblems. In \cite{yu2018uav}, the authors work on optimizing both the flight paths of UAVs and the distribution of radio resources to serve the maximum number of IoT devices. Authors successfully found optimal solutions for small-scale scenarios using the branch, reduce, and bound algorithm, while developed less-than-perfect solutions for larger scenarios.

In this paper, our primary focus is on the strategic placement of UAVs within a service area, where UAVs can serve as fog nodes to provide support to affected mobile user nodes on the ground. It is essential to underscore that the positioning of UAVs plays a pivotal role in various activities, including but not limited to rescue operations and exploration. To illustrate this importance, we considered a scenario in which a large number of individuals, impacted by a disaster, are equipped with wearable devices that continuously gather real-time health data. In such scenarios, two fundamental questions emerge. First, what is the minimum number of UAVs required? Second, where should UAVs be precisely located above the affected area to ensure optimal coverage for all affected persons? Determining the optimal placement of UAVs in this specific context is not just significant but also profoundly challenging. To address the challenge of optimizing UAV placement to enhance rescue services within the Internet of Things (IoT), we introduce a heuristic algorithm based on the hunting behavior of humpback whales, a species of baleen whales. This algorithm has been carefully crafted to strategically position UAVs over the service area in a manner that caters to the demands of mobile user devices. Our research also encompasses extensive simulations, which convincingly demonstrate that our approach can substantially reduce the necessary UAV deployment numbers while still ensuring effective communication coverage. The main contributions of this paper are outlined as follows:

\begin{itemize}
    \item We introduce a novel model centered around dynamic UAV-based fog deployment. It emphasizes the system's adaptability and operational efficacy within the afflicted areas. This model revolves around the deployment of fog nodes in a layered structure, catering to mobile users' needs through cloud components, UAV fog nodes, and the users themselves.
    \item We formulate and implement a novel optimization strategy for UAV fog node placement in dynamic networks during evacuation scenarios, with a primary focus on ensuring robust connectivity and maximal coverage for mobile users, while extending the network's lifespan.
    \item We employ the Adaptive Whale Optimization Algorithm (WOA) for fog node deployment in dynamic network. Its agility, rapid convergence, and low computational demands make it an ideal fit for high-pressure environments.
    \item We conduct a series of experiments using diverse benchmark scenarios related to the application of UAV fog nodes in providing network connectivity and coverage. The results consistently demonstrate that our presented algorithm outperforms as compare to existing methods, effectively addressing challenges such as limited connectivity in dynamic environments and potential node failures.
\end{itemize}

The rest of this paper is organized as follows: Section \ref{sec.2} provides a literature review of related works, Section \ref{sec.3} introduces a UAV-based fog deployment model, detailing its architecture, operational dynamics, key assumptions, coverage, connectivity criteria, and essential notations. Following this, Section \ref{sec.4} defines the main problem: optimizing UAV fog node placement for robust connectivity and extended lifespan during serving. Introduce a novel coverage and connectivity approach and employ a two-phase optimization approach using Adaptive Whale Optimization Algorithm (WOA) and Energy-Conscious Node Swapping algorithms. Section \ref{sec.5} presents the numerical simulation results. Finally, in Section \ref{sec.6}, we draw our main conclusions and discuss future work.

\section{Related work}
\label{sec.2}

To construct a robust network, researchers have studied the UAV deployment problem with different objectives, aiming to optimize aspects such as network connectivity, coverage, and energy consumption. Most studies consider these factors separately or jointly. For instance, the authors in \cite{oubbati2019leveraging} deployed a fleet of UAVs to monitor, detect, and localize affected area accidents, effectively notifying response rescue teams about their locations and the most optimal routes to reach them. In \cite{malandrino2019planning}, authors deployed multiple UAV nodes to cover the designated regions. This involves positioning the UAVs strategically to ensure optimal coverage. Based on the collected data, they utilized path planning algorithms to determine the most efficient routes. Consistently, in \cite{klaine2018distributed}, authors employed a reinforcement learning technique to optimize the placement of UAVs, enhancing the network coverage over affected users in a large-scale natural disaster. In \cite{mozaffari2016efficient} and \cite{mozaffari2015drone}, the authors aimed to minimize the number of UAVs needed for area coverage. They utilized an Air-to-Ground path loss model originally proposed by \cite{alhourani2014modeling} for an aerial wireless base station. However, a clear constraint arises from their assumption that all users are situated outdoors. This assumption limits the applicability of their methods, as real-world scenarios necessitate consideration for both indoor and outdoor users. In \cite{alhourani2014optimal, wei2017coverage}, the study focuses on determining the optimal UAV altitude to ensure maximum coverage. Furthermore, \cite{wei2019capacity} investigates capacity enhancement through strategic UAV placement in relaying networks, while \cite{wei2020capacity} explores this concept in the context of wireless sensor networks (WSNs), and \cite{na2020join} delves into the realm of OFDM networks. Furthermore, studies in \cite{cui2021decision} explore coverage maximization based on the altitude of UAVs within an emergency network. In \cite{zhao2019uav}, the analysis delves into the capacity and overall outage probability of fixed altitude UAVs serving as relays for on-scene available devices (OSAs) during emergency scenarios. The discourse shifts to optimized UAV placement for periodically gathering information across multi-zone disaster regions in \cite{peer2020multiuav}. Similarly, \cite{gupta2020optimal} conducts a study on optimizing UAV placement to maximize coverage for OSAs with varying quality-of-service (QoS) requirements. Additionally, \cite{gupta2020optimal} employs computational intelligence techniques to optimize placement for ensuring coverage and QoS within an emergency network. Dhelim et al. discussed the application of UAV in scene classification from Very High-Resolution Image using Transformers \cite{Chaib2022}, Unknown traffic identification \cite{Wei2022}, Blockchain application in vehicular environments that involves UAV \cite{aung2022blockchain}, and various other UAV applications related to image processing \cite{web3,SahraouiDhelim2016,Dhelim2020,ipunet,dhelim2022artificial,dhelim2022hybrid,dhelim2021social}

The above literature primarily concentrates on UAV network placement with only focus on coverage, neglecting the critical aspects of network connectivity and overall network lifespan. This is particularly significant as UAVs, being energy-limited, have constrained working time. However, some studies, like \cite{brust2016vbca}, present the Virtual Forces Clustering Algorithm (VBCA), drawing inspiration from molecular geometry's VSEPR model. VBCA strategically arranges UAVs in a clustered swarm, assigning a central UAV as the cluster-head to influence the overall network topology. This algorithm aims to maximize volume coverage while maintaining advanced connectivity within the clustered UAV swarm. Another study \cite{duan2021dynamic} introduces a mobility model optimized for both connectivity and coverage in Unmanned Aerial Vehicle (UAV) networks. The proposed model assesses the trade-off between connectivity and spatial coverage, where the spatial coverage, or areal coverage, is quantified as the percentage of the sensed target area within a specified time frame and connectivity is measured as the percentage of time UAVs remain connected to a sink, averaged across all UAVs. Furthermore, several algorithms, like particle swarm optimization, artificial bee colony, and ant colony optimization (ACO), have been suggested to control and coordinate groups of UAVs in various search, rescue, and tracking applications \cite{senanayake2016search, duan2021dynamic}. However, none of the previously mentioned studies has focused on optimizing UAV positions to simultaneously extend network coverage, enhance connectivity, and prolong network lifespan. To the best of our knowledge, this work is the first to address the tradeoff between coverage and connectivity in UAV placement while considering the energy limitations of the UAVs.

Table \ref{tab:literature-summary} presents the summary of the previous literature methods and their limitations to connectivity, coverage and energy which are our concern on this research.

\begin{table*}[htbp]
    \centering
    \caption{Summary of Previous Literature Methods and Limitations}
    \label{tab:literature-summary}
    \begin{tabular}{|c|p{5.5cm}|c|c|c|}
    \hline
    \textbf{Reference} & \textbf{Method} & \textbf{Connectivity} & \textbf{Coverage} & \textbf{Energy} \\
    \hline
    \cite{oubbati2019leveraging} & Monitoring, detecting, and localizing affected area accidents using UAVs for notifying rescue teams. & $\checkmark$ & $\times$ & $\times$ \\
    \cite{malandrino2019planning} & Strategic positioning of UAVs for optimal coverage, utilizing path planning algorithms for efficient routes. & $\checkmark$ & $\times$ & $\times$ \\
    \cite{klaine2018distributed} & Employing reinforcement learning to optimize UAV placement for enhanced network coverage during natural disasters. & $\checkmark$ & $\times$ & $\times$ \\
    \cite{mozaffari2016efficient}, \cite{mozaffari2015drone} & Minimization of required UAVs for area coverage, using Air-to-Ground path loss models. & $\checkmark$ & $\times$ & $\times$ \\
    \cite{alhourani2014optimal}, \cite{wei2017coverage} & Determining optimal UAV altitude for maximum coverage. & $\checkmark$ & $\times$ & $\times$ \\
    \cite{wei2019capacity} & Capacity enhancement through strategic UAV placement in relaying networks. & $\checkmark$ & $\times$ & $\times$ \\
    \cite{wei2020capacity} & Exploring capacity enhancement in wireless sensor networks (WSNs) through UAV placement. & $\checkmark$ & $\times$ & $\times$ \\
    \cite{na2020join} & Investigating optimized UAV placement in OFDM networks. & $\checkmark$ & $\times$ & $\times$ \\
    \cite{cui2021decision} & Maximizing coverage based on UAV altitude in an emergency network. & $\checkmark$ & $\times$ & $\times$ \\
    \cite{zhao2019uav} & Analyzing capacity and outage probability of fixed altitude UAVs serving as relays during emergencies. & $\checkmark$ & $\times$ & $\times$ \\
    \cite{peer2020multiuav} & Optimizing UAV placement for gathering information across multi-zone disaster regions. & $\checkmark$ & $\times$ & $\times$ \\
    \cite{gupta2020optimal} & Using computational intelligence for optimized placement to ensure coverage and QoS in emergency networks. & $\checkmark$ & $\times$ & $\times$ \\
    \cite{brust2016vbca} & Virtual Forces Clustering Algorithm (VBCA) for clustered UAV arrangement to maximize volume coverage while maintaining connectivity within the swarm    & $\checkmark$ & $\checkmark$ & $\times$ \\
    \cite{duan2021dynamic} & Mobility model optimizing connectivity and coverage by assessing the trade-off between spatial coverage and connectivity percentage. 
    & $\checkmark$ & $\checkmark$ & $\times$ \\
    \cite{senanayake2016search}, \cite{duan2021dynamic} & Algorithms like particle swarm optimization, artificial bee colony, and ant colony optimization for controlling and coordinating groups of UAVs in search, rescue, and tracking applications. & $\checkmark$ & $\checkmark$ & $\times$ \\
    Our work & Addresses the trade-off between coverage and connectivity in UAV placement while considering energy limitations, aiming to extend network coverage, enhance connectivity, and prolong network lifespan. & $\checkmark$ & $\checkmark$ & $\checkmark$ \\
    \hline
    \end{tabular}
\end{table*}

\section{System model}
\label{sec.3}

\begin{figure}[h]
    \centering
    \includegraphics[width=0.5\textwidth]{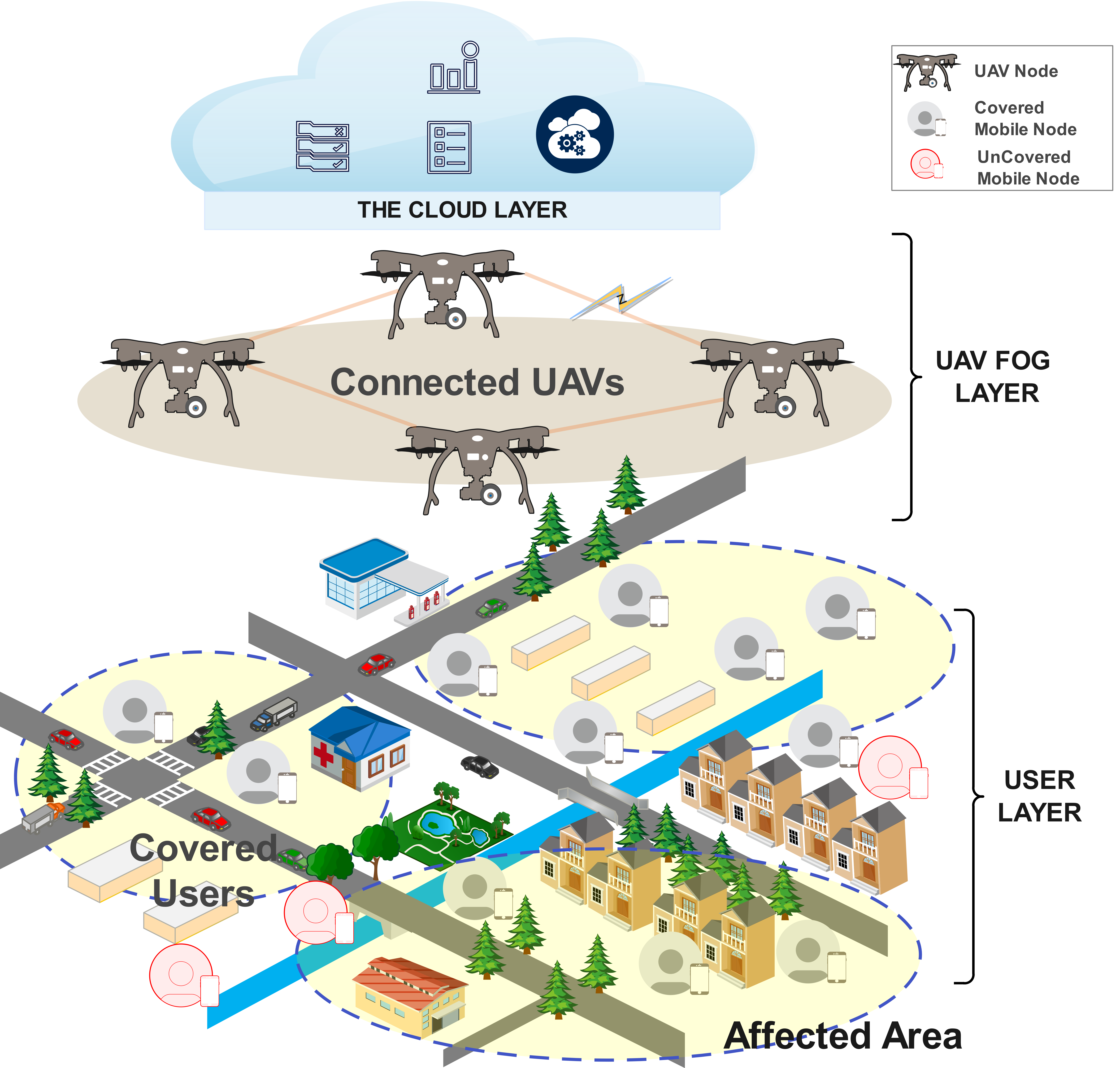}
    \caption{UAV-based fog infrastructure}
    \label{fig:uav-fog}
\end{figure}

This section introduces a dynamic UAV-based fog deployment model designed to operate within afflicted areas. The system we investigate encompasses mobile users within these afflicted regions and deploys UAVs as mobile fog nodes, as illustrated in Figure \ref{fig:uav-fog}. Our proposed UAV-based fog infrastructure is composed of three distinct layers. The uppermost layer features the cloud component, a centralized computing resource. In the middle layer, we deploy UAV fog nodes, capable of easily and intelligently moving through the afflicted area to provide fog computing capabilities where needed. Finally, the bottom layer consists of the mobile users, who are the primary beneficiaries of this dynamic fog deployment system. This multi-layered UAV fog nodes-based infrastructure enables us to create various fog computing environments that cater to the specific needs of users in afflicted areas. The dynamic nature of this infrastructure ensures that fog resources are efficiently allocated and dynamically adjusted in response to evolving conditions and user requirements, ultimately improving the overall resilience and performance of the system.

UAV fog nodes are not restricted to fixed positions like traditional base stations or cell towers. They can be deployed quickly and repositioned as needed to ensure maximum coverage and connectivity in real time. This agility enables them to adapt to changing user density and location. In situations such as natural disasters or emergency response scenarios, UAVs can be smoothly deployed to areas with disrupted or no communication infrastructure. They act as "flying base stations," offering immediate connectivity to affected users. They can provide a line-of-sight connection to users, which is especially useful in remote or challenging terrain. Also, they can identify areas with a high concentration of users and prioritize coverage in those regions, ensuring efficient use of resources. 

This study involves the deployment of a limited number of Aerial Vehicle (UAV) fog nodes in a three-dimensional geographical area. It is assumed that each UAV fog node can maintain constant communication with the central cloud infrastructure via cellular networks. Additionally, the positions of mobile users remain quasi-static. 

We define the set of nodes within the dynamic UAV-based fog Computing System (UFCS) as $V$, which can be represented as the union of two subsets: $F$ and $U$.
\begin{itemize}
    \item $F$ represents the set of Unmanned Aerial Vehicles (UAVs) fog nodes and is denoted as $F = \{f_1, f_2, \dots, f_n\}$, where each UAV node, labeled by $i$, possesses a radio coverage with a radius $\gamma_i$ and serves as a UAV mobile fog node capable of dynamic deployment across various network scenarios.
    \item $U$ represents the set of mobile affected users and is represented as $U = \{u_1, u_2, \dots, u_m\}$. Each user, labeled by $j$, is part of this set.
\end{itemize}

The UFCS operates in a dynamic environment where all UAV fog nodes exhibit mobility, allowing their positions to change. Similarly, mobile user devices can dynamically power on or off as needed, driven by factors like battery life or network conditions. The central challenge is to periodically adjust the positions of the mobile UAV fog nodes to ensure the UFCS network topology remains adaptable to these ever-changing conditions. To address this, the study divides time into discrete timeframes, with each timeframe signifying a specific period. During each timeframe ($\tau$-th), each user, denoted as $u_i \in U$, is placed at a two-dimensional coordinate point, $P_{\tau}(u_i) \in \mathbb{R}^2$, within the designated deployment area. The primary objective in serving these mobile users at the $\tau$-th timeframe is to determine the positions of UAV fog nodes, represented as $P_{\tau}(F) = \{P_{\tau}(f_1), P_{\tau}(f_2), \dots, P_{\tau}(f_n)\}$. Here, $P_{\tau}(f_i)$ signifies the spatial coordinates of each UAV fog node $(x_{\tau}, y_{\tau}, z_{\tau})$, i.e., $f_i$, in the deployment area at the $\tau$-th timeframe, where $i$ ranges from 1 to $n$.

In our scenario, each UAV fog node is equipped with a radio coverage range, which is a variable parameter influenced by various factors rather than a fixed attribute. These factors include the UAV's altitude, transmission power, antenna specifications, environmental conditions, and regulatory constraints. As in \cite{wang2023uav}\cite{lin2021adaptive}, for the sake of simplification within our model, we make the assumption that UAV fog nodes fly at the same latitude. This uniformity aids in modeling and analysis. This coverage range, denoted as $\gamma_{i\tau}$, which consists of an area centered at position $P_{\tau}(f_i)$ with a radius of $\gamma_i$. Our objective is to determine the optimal positions of these UAV fog nodes during each $\tau$-th timeframe.

To achieve this, the study establishes a network topology graph for every timeframe, known as $G_{\tau} = (V_{\tau}, E_{\tau})$. In this graph, $V_{\tau}$ is the set of all nodes, comprising $F$ (the UAV fog devices) and $U_{\tau}$ (the mobile user devices that are currently active). Notably, for each pair of UAV fog devices, $f_i$ and $f_j$ belonging to $F$, a UAV fog-fog link $(f_i, f_j)$ exists in $E_{\tau}$ if the radio coverage areas, $\gamma_{i\tau}$ and $\gamma_{j\tau}$, overlap or intersect. This signifies that they can communicate with each other. Additionally, for any mobile user device $u_j$ within $U_{\tau}$ and any UAV fog device $f_i$ within $F$, if the spatial position of $u_j$, $P_{\tau}(u_j)$, falls within the coverage range of $f_i$, i.e., $P_{\tau}(u_j) \in \gamma_{j\tau}$, a user-UAV fog link $(u_j, f_i)$ exists in $E_{\tau}$. This implies that the user device $u_j$ can communicate with the UAV fog node $f_i$.

\subsection{Underlying presumptions of the system model}

In our investigation of real-world network deployment scenarios, we examine a UAV fog deployment setting in which the fog infrastructure is dynamic, while the user nodes remain quasi-static. This dynamic environment is characterized by the homogeneity of UAV fog nodes where UAV fog nodes had the same communication radius. Furthermore, these UAV fog nodes have the capability to establish communication links with each other within the boundaries of their radio coverage radius, as described in prior research \cite{zhao2018deployment}. Additionally, Mobile user nodes are unable to communicate with each other directly, thus making it obligatory for their communication to traverse through UAV fog nodes in order to establish connections with other mobile user nodes. For the sake of simplifying the problem at hand, we summarize the following assumptions to establish optimal network connectivity and coverage:
\begin{itemize}
    \item Uniform and Random Deployment of Nodes: We assume that both the UAV fog nodes and mobile user nodes are deployed uniformly and in a random manner throughout the entire network \cite{wang2019adaptive}. Notably, the positions of these nodes are subject to changes, rendering them inherently dynamic within the predefined affected area.
    \item Access to Cloud Center through Cellular Networks: Each UAV fog node situated within the defined region has seamless access to the central cloud center via cellular networks.
    \item Uniform Communication Ranges for UAV Fog Nodes: We assume that all UAVs have the capability to operate at a consistent fixed altitude, ensuring uniform coverage \cite{zhao2018deployment,wang2019adaptive,8448770,8891520}. This assumption aligns with the common practice of standardizing communication capabilities among aerial devices in practical deployments.
    \item Limited Connections to Terminal User Nodes: We assume that a UAV fog node has the capacity to establish connections with only a limited number of user nodes. This constraint is essential for our analysis, as it aligns with the operational restrictions that UAV fog nodes often encounter in practical scenarios, where resource limitations may necessitate selective connections.
    \item Line-of-Sight Dominated Wireless Channel: Similar to \cite{azari2017ultra,wang2020packet}, we assume that the wireless channel between ground user $i$ and UAV is predominantly line-of-sight dominated. This assumption allows us to utilize the free space path loss model to characterize the wireless communication channel.
\end{itemize}

\begin{figure}
    \centering
    \includegraphics[width=0.5\textwidth]{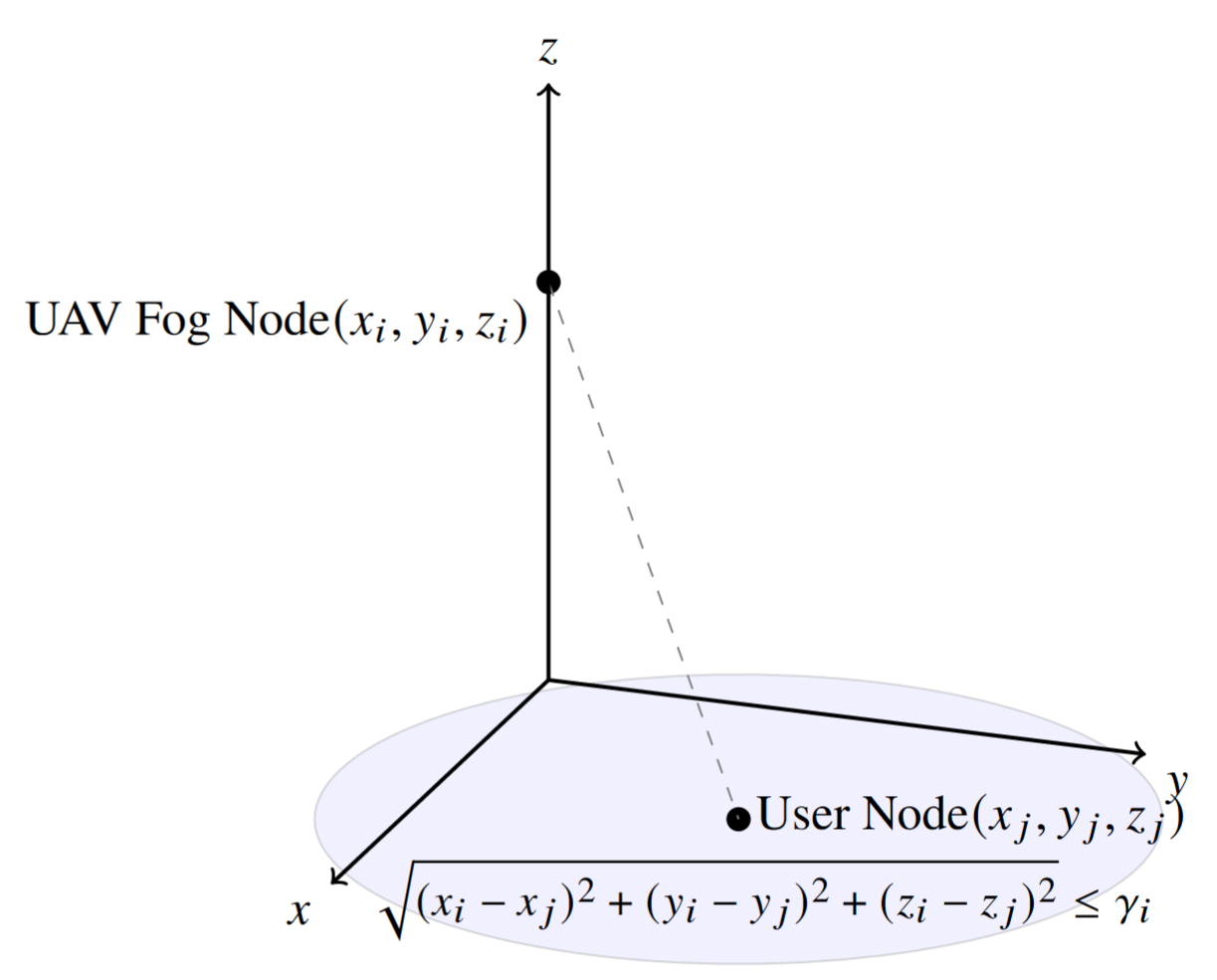}
    \caption{UAV and user nodes communication condition within 3D visualization}
  \label{fig:uav_network_equation_compact}
\end{figure}

To achieve the desired network performance, it is imperative to satisfy two essential conditions:
\begin{enumerate}
    \item A mobile user $u_i$ is regarded covered only when it is within the coverage range of at least one UAV fog node (Figure~\ref{fig:uav_network_equation_compact}), denoted as $f_j$, as demonstrated in Equation (\ref{eq:distance}):
    \begin{equation}\label{eq:distance}
    \sqrt{(x_i-x_j)^2 + (y_i-y_j)^2 + (z_i-z_j)^2} \leq \gamma_i
    \end{equation}
    The equation uses the Euclidean distance formula to determine if the distance between the user node $(x_i,y_i,z_i=0)$ and the fog node $(x_j,y_j,z_j)$ is less than or equal to the minimum coverage range between the user node $(\gamma_i)$ and the fog node $(\gamma_j)$. While the altitude $(z_i)$ is at ground level (0). The coverage condition is expressed as follows:
   \begin{equation}\label{eq:distance_H}
    \sqrt{(x_i-x_j)^2 + (y_i-y_j)^2 + H^2} \leq \gamma_i
   \end{equation}
\end{enumerate}

For two UAV fog nodes, $f_i$ and $f_j$, to be regarded as connected, it is essential that they fall within each other's communication range, as expressed in Equation (\ref{eq:distance_fog_user}):
\begin{equation}\label{eq:distance_fog_user}
    \sqrt{(x_i-x_j)^2 + (y_i-y_j)^2} \leq \gamma_i
\end{equation}

\begin{figure*}[!htbp]
	\centering
    \begin{subfigure}{0.4\textwidth}
        \centering
        \includegraphics[width=\linewidth]{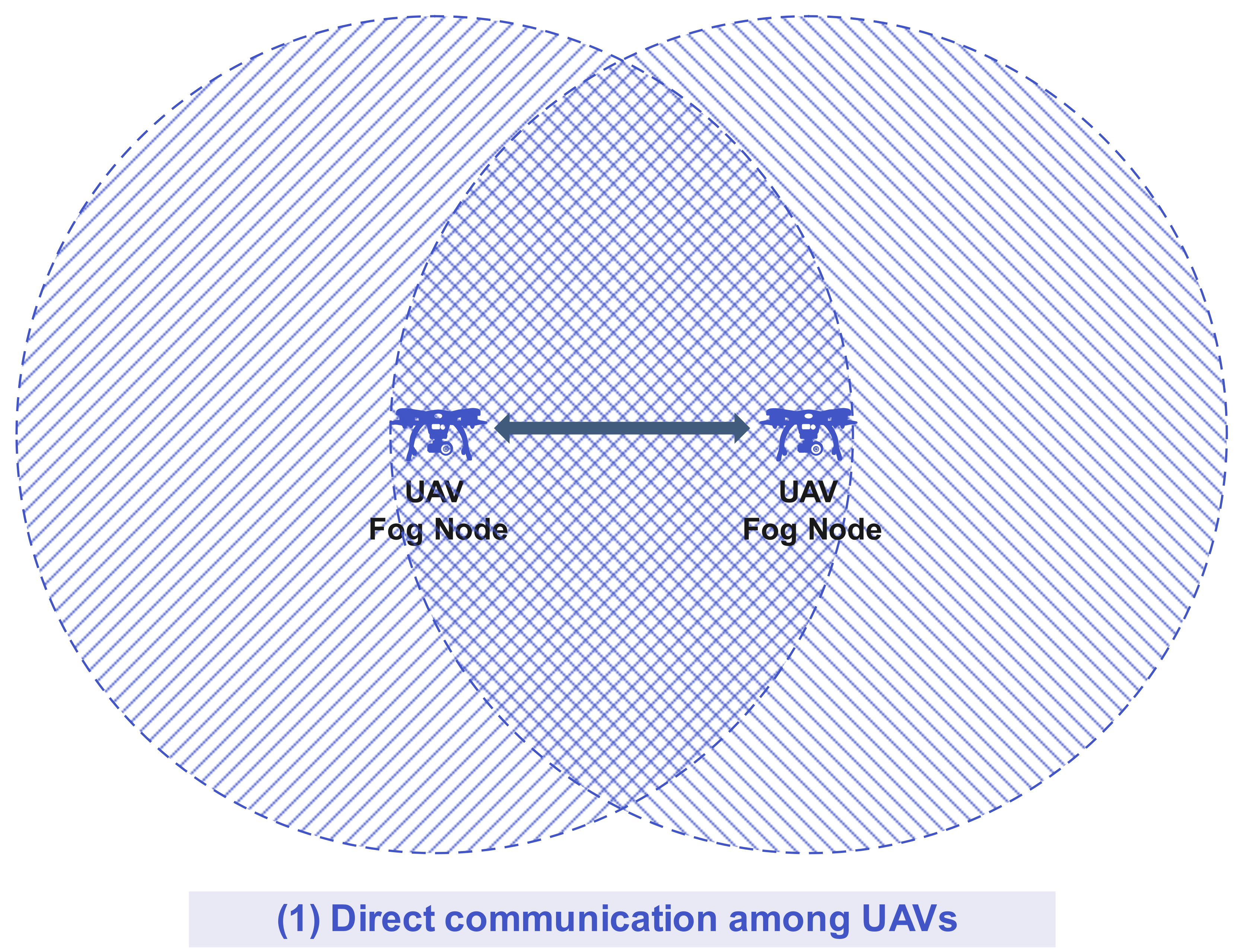}
        \caption{}
        \label{fig:sub1}
    \end{subfigure}
    \hfill
    \begin{subfigure}{0.46\textwidth}
        \centering
        \includegraphics[width=\linewidth]{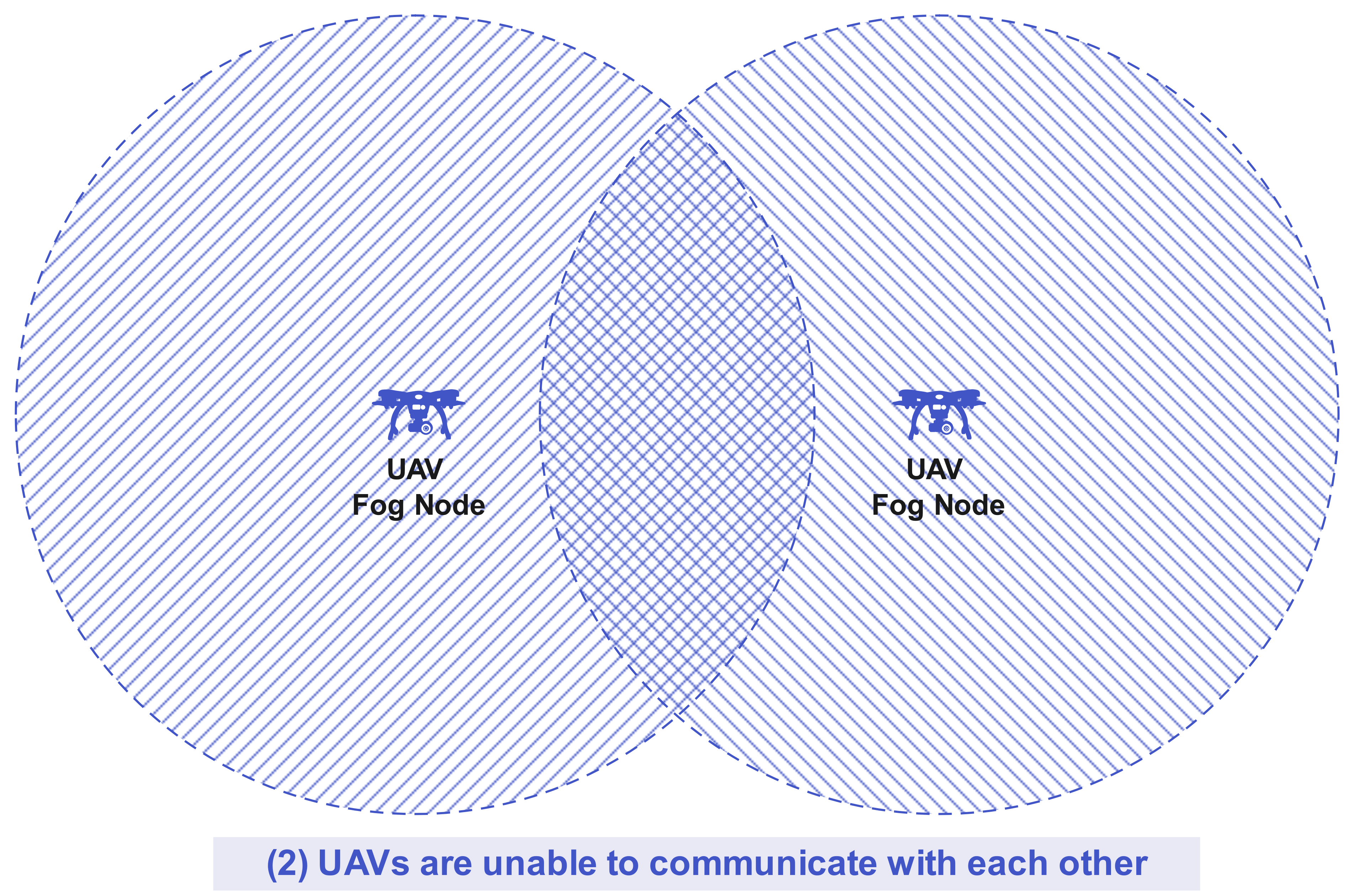}
        \caption{}
        \label{fig:sub2}
    \end{subfigure}
    \caption{UAVs communication condition within 2D visualization}
    \label{fig:uavs_communication_combined}
\end{figure*}

Figure~\ref{fig:uavs_communication_combined} illustrates the UAVs' communication condition within a 2D visualization.

To facilitate comprehension and reference, we provide a summary of the primary notations used throughout this paper in \ref{tab:key_notations}.

\begin{table}[h]
    \centering
    \caption{Key Notations Description}
    \label{tab:key_notations}
    \resizebox{\linewidth}{!}{%
        \begin{tabular}{|>{\hspace{0pt}}p{0.25\linewidth}|>{\hspace{0pt}}p{0.85\linewidth}|} 
            \hline
            Symbol & Meaning \\ 
            \hline
            $n$ & Total count of UAV fog Nodes. \\
            \hline
            $m$ & Total count of end-users within the affected area. \\
            \hline
            $F$ & Unmanned Aerial Vehicles (UAVs) fog nodes. \\
            \hline
            $U$ & Set of mobile affected end-users. \\
            \hline
            $V= F\cup U$ & Both the UAV fog nodes ($F$) and the mobile affected users ($U$) within the UAV-based Fog Computing System (UFCS). \\
            \hline
            $P$ & Position of UAV Fog Node. \\
            \hline
            $\tau$ & Timeframe. \\
            \hline
            $P_{\tau}(fi)$ & Locations of each UAV fog node in space at the $\tau$-th timeframe. \\
            \hline
            $\gamma_i^\tau$ & Radio coverage radius of $fi$ at $\tau$-th timeframe. \\
            \hline
            $G^\tau=(V^\tau, E^\tau)$ & The topology graph for every timeframe $\tau$. \\
            \hline
            $E^\tau$ & Interconnecting edges between UAV fog nodes at timeframe $\tau$. \\
            \hline
            $P_j$ & The power of UAV node $j$. \\
            \hline
            $G_i$ & Isolated subgraph component $G_i$ at index $i$. \\
            \hline
            $|G_i|$ & Number of elements comprising $G_i$. \\
            \hline
            $C_1^i$ & Binary variable indicates coverage condition over all network. \\
            \hline
            $C_2^i$ & Binary variable indicates coverage condition within the largest network. \\
            \hline
            $\text{NCV}_2^{ }(G^*)$ & Extent of coverage for affected end-user nodes. \\
            \hline
            $\text{ENG}_i(G^*)$ & Extent the energy of UAV fog node $i$. \\
            \hline
            $\text{NLS}(G^*)$ & Life Time of the network $G^*$. \\
            \hline
            $\text{ENG}_{(j)}^t$ & Energy consumed during hovering and traveling at time $t$. \\
            \hline
            $\text{ENG}_{ht}^{(t)}$ & Energy consumed during hovering and traveling at $t$. \\
            \hline
            $\text{ENG}_c^t$ & Energy consumed during transmitting and receiving at $t$. \\
            \hline
        \end{tabular}
    }
\end{table}

\section{UAV Fog Node Placement: Problem Definition and Optimization}
\label{sec.4}

Our central objective revolves around the determination of an optimal placement strategy for UAV fog nodes during evacuation with extending the network lifespan. This involves carefully adjusting the positions of UAV fog nodes within the affected area, aiming to ensure robust connectivity, maximize coverage and extending network lifespan.


\subsection{Problem Definition}
Given the complex and dynamic nature of real-world scenarios, attempting to analyze the entire network is often impractical, if not infeasible, primarily due to potential network disconnectivity and deployment cost where we could not provide UAV to cover all the affected area. To address this, we narrow our focus to a more manageable challenge: identifying the most connected sub-network, often referred to as the primarily connected sub-network. This approach allows us to work within the confines of practical constraints and connectivity limitations. Let's consider a graph $G$ that is composed of $k$ subgraphs, denoted as $G_1, \ldots, G_k$ within $G$. In this context, $G$ is the union of these subgraphs, which can be expressed as $G = G_1 \cup G_2 \cup \ldots \cup G_k$. It's crucial to emphasize that the intersection of any two subgraphs, $G_i$ and $G_j$, where $i$ and $j$ belong to set $1, \ldots, k$, is empty. This means that $G_i \cap G_j = \emptyset$ for all $i$ and $j$ within the range of $1$ to $k$.

In order to select the most connected network of Graph $G$, we determine the connectivity of each subgraph, $G_1$ through $G_k$, by evaluating the connections or links between UAV nodes within each subgraph as expressed in Equation (\ref{eq:NC_max_G}):

\begin{equation}\label{eq:NC_max_G}
    \text{NC}(G^*) = \arg\max_{i \in \{1,2,\ldots,k\}} |G_i|
\end{equation}

To model the network coverage, it's important to define the coverage condition. In the context of UAV fog nodes and mobile users, coverage typically refers to the condition where a mobile user is considered to be within the communication range of at least one UAV fog node \cite{context_siot,BenSada2023}. To our current knowledge, prior studies \cite{lin2020dynamic,taleb2022solving} have traditionally defined client coverage in the context of network graphs as follows:
\begin{equation}\label{eq:NCV_1_G}
    \text{NCV}_1(G) = \sum_{i=0}^m C_1^i
\end{equation}
Where
\begin{equation}\label{eq:C_1^i}
    C_1^i = \begin{cases} 
                1, & \text{if } \text{distance}(P_i,F_j) \leq R \text{ for at least one } F_j \text{ in } G \\
                0, & \text{otherwise}
            \end{cases}
\end{equation}
Where $C_1^i$ equals $1$ if any UAV fog node covers end-user $i$ within the whole network graph $G$, and $0$ otherwise. In this approach, balancing mobile user coverage and network connectivity has been acknowledged as conflicting objectives.

However, in our revised formulation prioritizing network connectivity, we redefine mobile user coverage by focusing only on end-users covered within the largest connected sub-graph, denoted as $G^*$. Hence, in mathematical terms, the coverage metric is expressed as follows:
\begin{equation}\label{eq:C_2^i}
    C_2^i = \begin{cases} 
                1, & \text{if } \text{distance}(P_i,F_j) \leq R \text{ for at least one } F_j \text{ in } G^* \\
                0, & \text{otherwise}
            \end{cases}
\end{equation}
This condition ensures that the mobile user can establish a network connection and receive network services. Hence, the network coverage can be expressed as:
\begin{equation}\label{eq:NCV_2_G_star}
    \text{NCV}_2(G^*) = \sum_{i=0}^m C_2^i
\end{equation}
Here, $C_2^i$ equals $1$ if any UAV fog nodes cover mobile user $i$ within the $G^*$ sub-graph (the largest connected component), and $0$ otherwise. This refined approach ensures that end-user coverage aligns with maintaining the robustness of the network's connectivity.

The coverage ratio variation across the affected area significantly impacts the network's ability to sustain operations. For instance, following a natural disaster, the communication demand in the affected area significantly rises, leading to a substantial increase in UAV energy consumption. This escalation in energy usage consequently shortens the network's lifespan. Therefore, the maintenance of an active and interconnected network becomes pivotal for successful disaster relief efforts. Thus, to ensure an extended operational period for the network, the primary focus is on maximizing its lifespan. We have introduced an energy metric (ENGi) for each UAV fog node (\(F_i\)), directly influenced by the density of mobile users covered by \(F_i\). It's evident that UAV nodes with higher coverage ratios consume more energy than those with lower coverage ratios. Therefore, to extend the network's lifetime, we have defined an energy metric (ENGi) for each UAV fog node (\(F_i\)), where \(ENG_i\) is directly affected by the number of mobile users covered by \(F_i\). This implies that UAV fog nodes with higher energy levels can accommodate a high number of users, while those with lower energy levels can cover fewer users. Hence, the total network lifespan (\(NLS\)) can be expressed as:

\begin{equation}\label{eq:NLS_G_star}
    NLS(G^*) = \sum_{j=0}^{|G_t^*|} ENG_j
\end{equation}

To facilitate calculations in assessing energy dynamics within communication systems, primarily, we presumed that all data packets maintain uniform sizes throughout their transmission as indicated \cite{lin2021adaptive}. Additionally, under specified conditions of meeting minimum power requisites within designated time frames, the presumption stands that all data packets can be reliably transmitted and received. These assumptions underscore the integral relationship between energy consumption in communication processes and both the size of transmitted data and the power essential for effective transmission and reception.

The energy consumption of a UAV fog node \(j\) during its service is expressed as follows (Equation \ref{eq:ENG_j_t}):
\begin{equation}\label{eq:ENG_j_t}
    ENG_{j}^{t}= ENG_{ht}^{t} + ENG_{c}^{t}
\end{equation}
Where \(ENG_{ht}^{t}\) refers to the energy consumed during hovering and traveling \cite{8908666}, and \(ENG_{c}^{t}\) refers to the communication usage during the transmitting and receiving signals involving covered mobile users, data transmission rate, and channel gain.

Given that the UAV maintains a steady altitude, its gravitational potential energy remains constant. In this simplified flight model that focuses solely on traveling and hovering energy, hence, the total energy consumption \(ENG_{ht}^{t}\) of the UAV at time \(t\) can be expressed as the sum of energy consumed during travel \(ENG_{travel}^{t}\) and energy consumed during hovering:
\begin{equation}\label{eq:ENG_ht_t}
    ENG_{ht}^{t}= ENG_{travel}^{t} + ENG_{hovering}^{t}
\end{equation}

The energy consumed during travel (\(ENG_{travel}^{t}\)) primarily depends on the UAV's flight distance and flight speed. This can be calculated using:
\begin{equation}\label{eq:ENG_travel_t}
    ENG_{travel}^{t}= P_{traveling} \times T_{travel}
\end{equation}
Where \(P_{traveling}\) represents the power required for traveling, and \(T_{travel}\) denotes the time duration of travel.

The energy consumed during hovering (\(ENG_{hovering}^{t}\)) is primarily determined by the UAV's hovering power and the duration of time it spends in a hovering state:
\begin{equation}\label{eq:ENG_hovering_t}
    ENG_{hovering}^{t}= P_{hovering} \times T_{hovering}
\end{equation}
Where \(P_{hovering}\) represents the power required for hovering, and \(T_{hovering}\) denotes the time duration spent in the hovering state.

\begin{figure*}[t]
    \centering
    \includegraphics[width=0.65\linewidth]{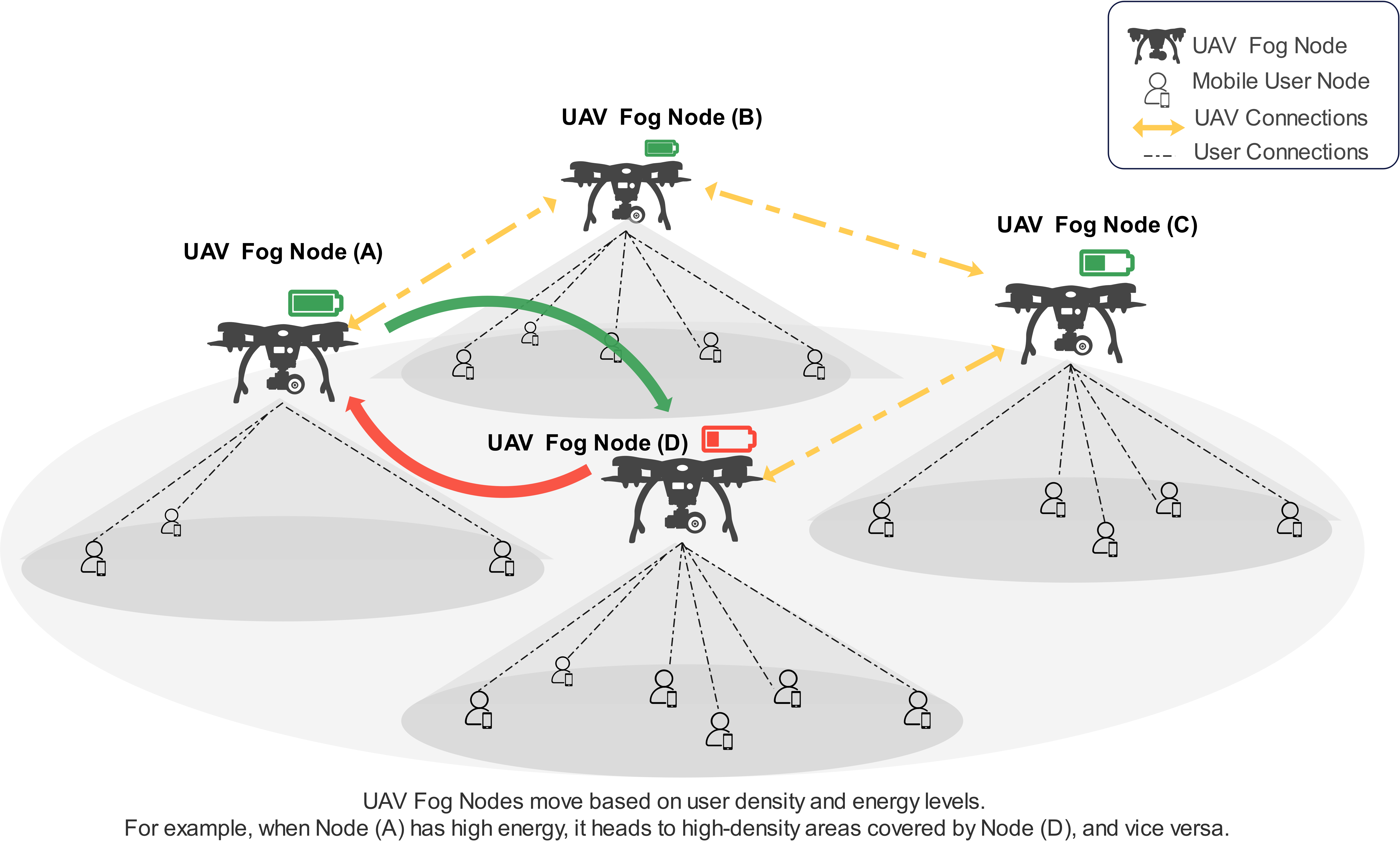}
    \caption{Optimizing UAV Fog Node Energy Consumption for Disaster Communication}
    \label{fig:figure4}
\end{figure*}

In order to evaluate the $ \text{ENG}_{c}^{t} $, As in \cite{8908666}, we assumed that each user device is limited to using only one channel to communicate with the UAV. This limitation ensures that each device operates within a specific communication frequency range without overlapping with others, and the channel state, which represents the quality of communication between the IoT devices and the UAV, remains constant during each timeframe. 

To estimate the quality of the communication channel (channel gains) between UAV fog node and the user devices, we adopted the air-to-ground channel model where the communication link between the UAV fog node and mobile users is dominated by the LoS path. The channel gain between the UAV fog node $ f_{j} $ and user $ u_{i} $ can be derived as:

\begin{equation}\label{eq:g_f_u}
    g(f_{j},u_{i}) = \frac{\beta_{0}}{((x_{i}-x_{j})^{2} + (y_{i}-y_{j})^{2} + H^{2})}
\end{equation}

The parameter $ \beta_{0} $ represents the channel gain at a reference distance of 1 meter.

Additionally, the channel gains for uplink (from IoT devices to the UAV) and downlink (from the UAV to IoT devices) considered reciprocal, for simplifying the communication process. Hence, we can calculate the energy consumption required to transmit data as follows:
    
\begin{equation}\label{eq:ENG_c_j_t}
    \text{ENG}_{c,j}^{t} = \sum_{i=1}^{N_{\text{users}}} ( \text{Energy}_{e_{i}} + \text{Energy}_{r_{i}} ) \times \text{Time}
\end{equation}

Where $ \text{Energy}_{e_{i}} $, $ \text{Energy}_{r_{i}} $ expressed as follow: 

\begin{equation}\label{eq:Energy_e_i}
    \text{Energy}_{e_{i}} = P_{e} \times \frac{\text{InputData}}{R_{(i,j)}^{\text{uplink}}}
\end{equation}

\begin{equation}\label{eq:Energy_r_i}
    \text{Energy}_{r_{i}} = P_{r} \times \frac{\text{OutputData}}{R_{(i,j)}^{\text{downlink}}}
\end{equation}

$ \text{Energy}_e $ is the energy consumed by the UAV node for each uplink transmission between the UAV $ f_{j} $ and IoT user $ u_{i} $. $ \text{Energy}_r $ is the energy consumed by the UAV node for each downlink transmission between the UAV $ f_{j} $ and the IoT user $ u_{i} $. While $ \frac{\text{InputData}}{R_{(i,j)}^{\text{uplink}}} $ and $ \frac{\text{OutputData}}{R_{(i,j)}^{\text{downlink}}} $ represents the duration of transmission, respectively.

Here, $ P_{e} $ and $ P_{r} $ represent the transmission and receiving power, respectively. $ R_{ij} $ denotes the data rate between user $ u_{i} $ and the UAV fog node $ j $. This data rate ($ R_{ij} $) can be formulated by:

\begin{equation}\label{eq:R_uplink}
    R_{(i,j)}^{\text{uplink}} = B \log_{2} \left(1 + \frac{P_{u}(i) \cdot g(f_{i},u_{i})}{\sigma^{2}}\right)
\end{equation}

\begin{equation}\label{eq:R_uplink_ij}
    R_{(i,j)}^{\text{uplink}} = B \log_{2} \left(1 + \frac{\beta_{0} P_{e}(j)}{\sigma^{2} ((x_{i}-x_{j})^{2} + (y_{i}-y_{j})^{2} + H^{2})}\right)
\end{equation}

\begin{equation}\label{eq:R_downlink_ij}
    R_{(i,j)}^{\text{downlink}} = B \log_{2} \left(1 + \frac{\beta_{0} P_{f}(j)}{\sigma^{2} ((x_{i}-x_{j})^{2} + (y_{i}-y_{j})^{2} + H^{2})}\right)
\end{equation}


In this equation, $ B $ stands for bandwidth, $ \beta_{0} $ represents the channel gain at a reference distance of 1 meter, $ P_{u}(i) $ represents the transmission power of the user node $ U_{i} $ and $ P_{f}(j) $ is the transmission power of the UAV node $ f_{j} $. $ \sigma^{2} $ denotes the noise power, and $ (x_{i},y_{i}) $ and $ (x_{j},y_{j}) $ are the coordinates of user $ u_{i} $ and the UAV node $ j $ respectively, in the spatial domain. The term $ H $ signifies the altitude or height difference between the devices and the correspond UAV fog node.

\subsection{Optimization Objectives:}

Our approach involves a two-phase strategy for optimizing the UAV fog node placement within a dynamic network topology. Initially, the optimization process incorporates two key objectives for optimization. The first objective is ensuring optimal network connectivity by determining the size of the largest subgraph component, denoted as $NC(G^*)$. The second objective is to improve mobile user node coverage, represented as $NCV_2^{ }(G^*)$ as outlined in equations \ref{eq:NC_max_G} and \ref{eq:C_2^i} respectively. Unlike previous studies \cite{lin2020dynamic} that relied on a weighted aggregated function, In our novel approach, we prioritize maximizing mobile user coverage within the most significant sub-graph component of the network to prevent network fragmentation. Our specific objective function, denoted as $H(X)$, aims at achieving this goal, as expressed in equation (\ref{eq:H_X}), where $G^*$ represents the largest sub-graph component of the network.
\begin{equation}\label{eq:H_X}
    H(X)=\frac{\text{NCV}_2(G^*)}{m}
\end{equation}

It is critical to underscore that our novel objective function is not singularly focused; instead, it aims to simultaneously optimize both mobile user coverage and UAV nodes connectivity,

Following this initial optimization phase, a sophisticated post-optimization energy-conscious rearrangement of UAV fog nodes based on their energy levels to maximize efficiency is executed to fine-tune the deployment based on energy efficiency. The optimization process initiates an accurate repositioning of UAV fog nodes to leverage their energy capacities effectively. Where High-energy nodes are strategically relocated to densely populated areas, optimizing their potential to serve a larger number of users. Conversely, low-energy nodes are reassigned to sparser regions, aligning with the objective of maximizing energy conservation in less demanding areas. This approach combines an energy-conscious rearrangement process after optimization. It aims not just to boost network performance but also to adjust where fog nodes are placed to fit energy limits.

Aligned with the defined problem scopes, the following subsection introduces the Adaptive Whale Optimization Algorithm (WOA) to achieve optimal coverage of mobile users while concurrently maximizing UAV fog nodes connectivity, as outlined in Algorithm \ref{algo:woa}. Additionally, we introduce the Energy-Conscious Node Swapping Algorithm (ECNSA), specifically to extend the network lifespan. Its key steps are outlined in Algorithm \ref{algo:ECNSA}.

\subsection{Whale Optimization Algorithm (WOA)}

The Whale Optimization Algorithm (WOA) is a technique developed by authors in \cite{mirjalili2016whale} in 2016, inspired by the hunting behavior of humpback whales. This optimization method seeks to mimic how humpback whales encircle and track their prey. The hunting process of humpback whales involves two primary stages: exploration and exploitation. Similarly, WOA algorithm operates in two main phases to explore and exploit the search space for optimal solutions. During the exploration stage, the algorithm treats the current best solution found as the target prey. Since the best solution is initially unknown, the algorithm assumes the current best candidate solution to be the target and guides the other "whales" (representing potential solutions) to update their positions in the search space based on this assumed target. 

\begin{figure}[h]
  \centering
 
  \includegraphics[width=0.6\linewidth]{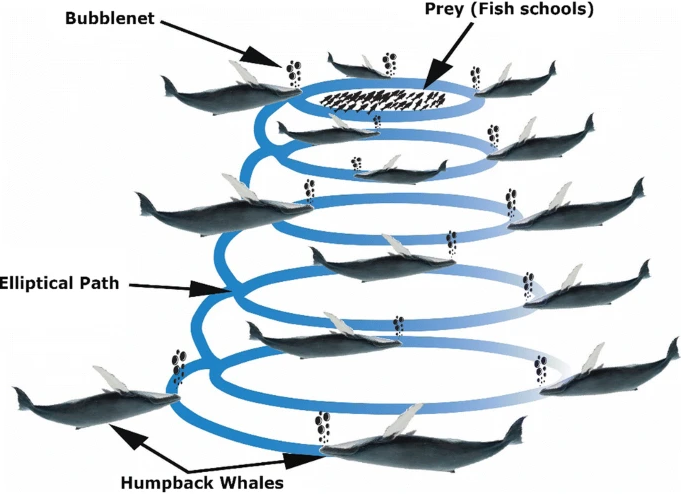}
   \label{fig:figure5}
    \caption{Bubble-net feeding behavior of humpback whales.}
\end{figure}

The mathematical modeling of this behavior is represented by equations \ref{eq:D} and \ref{eq:X_t+1} in the context of the WOA approach. These equations likely describe how the positions of solutions (UAV fog nodes coordinate) or "whales" are updated in relation to the assumed target solution, reflecting the algorithm's process of exploration and exploitation to find an optimal solution in the search space (optimal UAVs placement).

\begin{align}
\label{eq:D}
\vec{D} &= \left|\vec{C} \cdot (\vec{X}^*) (t) - \vec{X} (t)\right| \\
\label{eq:X_t+1}
\vec{X} (t+1) &= (\vec{X}^*) (t) - \vec{A} \cdot \vec{D}
\end{align}

Where $\vec{D}$ calculates the difference between the product of $\vec{C}$ and the best solution obtained so far ($\vec{X}^*$) at time $t$, and the current solution ($\vec{X} (t)$). And $\vec{X} (t+1)$ updates the current solution based on the difference $\vec{D}$, modifying it in relation to the best solution ($\vec{X}^*$) at time $t$ using coefficient vectors $\vec{A}$ and $\vec{C}$.

$\vec{A}$ and $\vec{C}$ are determined using specific equations \ref{eq:A_vec} and \ref{eq:C_vec}, involving vectors $\vec{a}$ and $\vec{r}$. These equations define how the coefficients $\vec{A}$ and $\vec{C}$ are computed.

\begin{align}
\label{eq:A_vec}
\vec{A} &= 2\vec{a} \cdot \vec{r} - \vec{a} \\
\label{eq:C_vec}
\vec{C} &= 2 \cdot \vec{r}
\end{align}

The variable $\vec{a}$ is a coefficient vector that linearly decreases from 2 to 0 over the range of iterations. This declining value of $\vec{a}$ plays a crucial role in controlling the balance between exploration and exploitation phases within the optimization process. Additionally, $\vec{r}$ represents a random vector with values ranging from 0 to 1. This random vector $\vec{r}$ plays a significant role in the Whale Optimization Algorithm by introducing randomness or stochasticity. This randomness enables each search agent which represented as a solution in the search space to explore various positions within the search space. This exploration aims to ensure that the algorithm doesn't get stuck in local optima and instead explores a wider range of possibilities. Therefore, the interplay between the linearly decreasing coefficient vector $\vec{a}$ and the random vector $\vec{r}$ contributes significantly to both the exploration and exploitation phases of the optimization process. The declining $\vec{a}$ values control the balance between these phases, while the random vector $\vec{r}$ introduces stochasticity to aid in exploration across the search space.

\begin{figure}[h]
  \centering
  \includegraphics[width=0.6\linewidth]{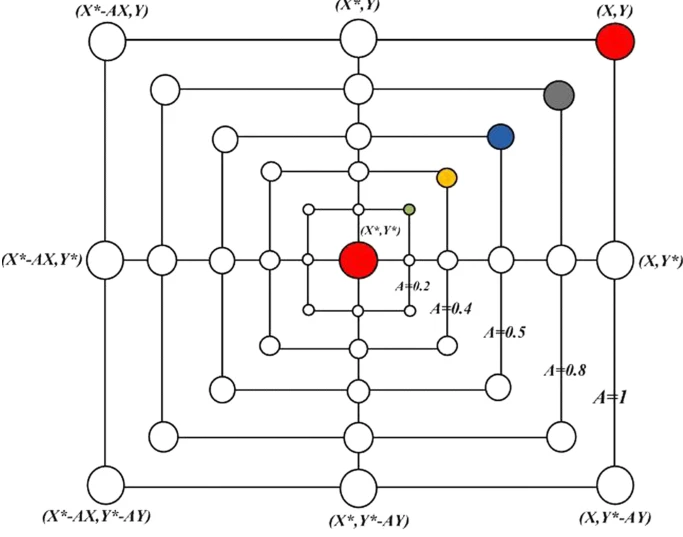}
   \caption{Bubble-net search mechanism ($X^*$ is the best solution obtained so far)}
   \label{fig:Shrinking_figure6}
\end{figure}

Within the WOA approach, two distinct foraging mechanisms, inspired by the strategies employed by humpback whales, are integrated to guide the search for optimal solutions:

\textbf{Shrinking Encircling Mechanism:}
    Mimicking the bubblenet technique used by humpback whales, this approach represents the exploitation strategy in the WOA. The algorithm minimizes the coefficient vector $\vec{A}$ by reducing the value of $\vec{a}$ (Equation \ref{eq:A_vec}). By doing so, search agents update their positions towards a new location, balancing between their original position and the current best solution. This technique mirrors the process of strategically encircling and narrowing down around a promising solution. Figure \ref{fig:Shrinking_figure6} shows a graphical representation of potential positions of a search agent, transitioning from the initial coordinates (X, Y) to a new position (X', Y'). This visual representation likely demonstrates how the agent's position changes over time or iterations within the defined parameter space, influenced by the shrinking $\vec{a}$ and the random values of $\vec{A}$.


\begin{figure}[h]
  \centering
  
  \includegraphics[width=0.6\linewidth]{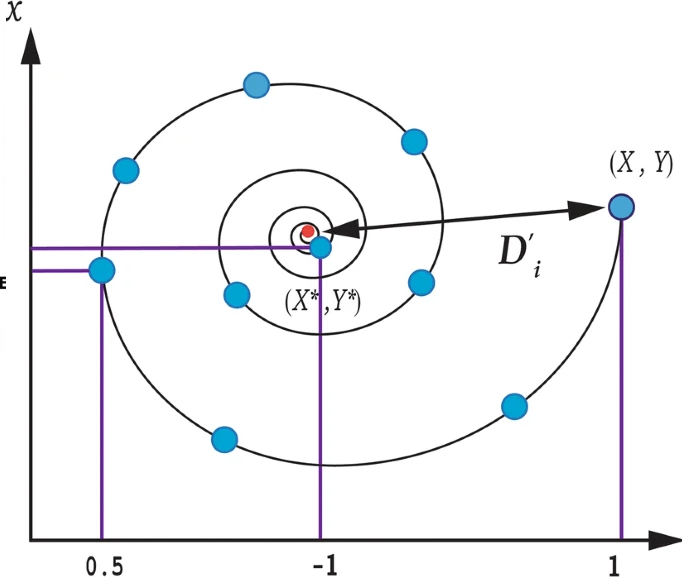}
  
  \caption{Spiral updating position}
  \label{fig:Spiral_updating_figure7}
 \end{figure}

\textbf{Spiral Updating Position:}
    To emulate the spiral-shaped motion observed in humpback whales during foraging, the algorithm computes the distance ($\vec{D}$) between the whale's and the prey's location (Equation \ref{eq:xdb_vector}). Using a spiral equation involving parameters such as $b$ controlling the spiral's shape and a random number $l$, the agent's position is updated along a trajectory that follows a spiral pattern between its current location and the location of the best solution found so far. This approach simulates the whale's spiral movement to navigate through the search space efficiently. Figure \ref{fig:Spiral_updating_figure7} illustrates how this combined approach affects the whale's movement around the prey. It shows how the position of the whale changes over time or iterations, influenced by the probabilistic choice between the two updating methods: the shrinking circle and the spiral movement.

    \begin{equation}
    \label{eq:xdb_vector}
    \vec{X}(t+1)=\vec{D}\cdot e^{b1}\cdot \cos(2\pi l)+\vec{X}^*(t)
    \end{equation}

\begin{algorithm}
    \SetAlgoLined
    \SetKwInOut{Input}{Input}
    \SetKwInOut{Output}{Output}
    
    \Input{Whales population $X_i(i=1,2, \ldots, n)$}
    \Output{Best search agent $X^*$}
    
    Initialize whales population\;
    Calculate fitness of each search agent\;
    $X^*=$ the best search agent\;
    
    \While{$(t<$ maximum number of iterations)}{
        \For{each search agent}{
            Update $a, A, C, l$, and $p$\;
            \If{$(p<0.5)$}{
                \If{$2(|A|<1)$}{
                    Update position of the current search agent by Eq. \ref{eq:D}\;
                }
                \ElseIf{$2(|A| \geq I)$}{
                    Select a random search agent $X_{\text{rand}}$\;
                    Update position of the current search agent by Eq. \ref{eq:D_vector}\;
                }
            }
            \ElseIf{$(p \geq 0.5)$}{
                Update position of the current search by Eq. \ref{eq:xdb_vector}\;
            }
        }
        Check if any search agent goes beyond the search space and amend it\;
        Calculate fitness of each search agent\;
        Update $X^*$ if there is a better solution\;
        $t=t+1$\;
    }
    \Return{$X^*$}
    
    \caption{Whale Optimization Algorithm}
    \label{algo:woa}
\end{algorithm}

Equation (\ref{eq:update}) dynamically chooses between two strategies, the Shrinking Encircling Mechanism and the Spiral Updating Position, using probabilistic selection and is structured as follows:
\begin{equation}
\label{eq:update}
\vec{X}(t+1)= \begin{cases} 
      \vec{X}^*(t)-\vec{A}\cdot\vec{D}" \text{ if } " p<0.5 \\
      \vec{D}\cdot e^{b1}\cdot\cos(2\pi l)+\vec{X}^*(t)" \text{ if } " p\geq0.5 
   \end{cases}
\end{equation}

The vector $\vec{A}$ plays a pivotal role in determining the exploration phase within the Whale Optimization Algorithm (WOA). When the magnitude of $\vec{A}$ exceeds 1, it serves as an indicator that the algorithm is predominantly in the exploration phase. During this exploration-dominant phase, the update of locations for the other search agents is primarily influenced by the best solution obtained thus far.

Consequently, the search agents prioritize updating their positions based on information derived from the current best solution, directing their exploration efforts within the search space. Therefore, the threshold of $\vec{A}$ acts as a critical criterion for the algorithm, delineating the exploration phase. It ensures that during phases where $\vec{A}$ surpasses this threshold, the focus remains on exploration, guiding the search agents to explore the search space by primarily leveraging information from the current best solution. 

Equations \ref{eq:D_vector} and \ref{eq:X_t+1_vector} constitute a fundamental mathematical model embodying an exploration strategy within the Whale Optimization Algorithm (WOA):
\begin{align}
\label{eq:D_vector}
\vec{D} &= |\vec{C}\cdot(\vec{X}_{\text{rand}})-\vec{X}| \\
\label{eq:X_t+1_vector}
\vec{X}(t+1) &= (\vec{X}_{\text{rand}})-\vec{A}\cdot\vec{D}
\end{align}

Where $\vec{X}_{\text{rand}}$ denotes a randomly chosen solution from the current pool of potential solutions.

Moreover, to further elevate the adaptability and responsiveness of the Whale Optimization Algorithm (WOA) across our problem, we introduced an adaptive exploration-exploitation mechanism. This mechanism dynamically adjusts the balance between exploration and exploitation phases based on the algorithm's performance during the execution, allowing for real-time adaptation to the problem characteristics. It aims to regulate key parameters such as the exploration coefficient vector $\vec{a}$ and the threshold $\vec{A}$, continuously tuning these values during the optimization process.

During initial iterations or when encountering exploration-dominant phases, the mechanism emphasizes exploration by promoting larger values of the exploration coefficient vector $\vec{a}$. This encourages wider exploration across the search space, preventing premature convergence and enabling the algorithm to explore uncharted regions efficiently. Conversely, as the algorithm progresses and approaches potential optima or convergence, the mechanism shifts focus towards exploitation by scaling down the exploration coefficient $\vec{a}$. This adaptation facilitates a more refined exploitation of promising regions, allowing the algorithm to converge towards optimal solutions effectively.


    
    
    
    

\subsection{Energy-Conscious Node Swapping Algorithm}

In the context of post-optimization for energy-conscious node repositioning, after an initial optimization phase achieving optimal coverage and connectivity, ECNSA can be employed to fine-tune the placement of UAV fog nodes based on energy considerations and population density \cite{fog_placement,Abdenacer2023}. Hence, in order to maximize the network lifespan. We design an optimization function as outlined in (\ref{eq:NLS_maximize}), that aims to maximize the overall network lifespan (NLS) by adjusting the obtained positions of UAV fog nodes based on their energy levels and coverage ratio. 

\begin{equation}\label{eq:NLS_maximize}
    \text{Maximize } NLS(G^*) = \sum_{i=0}^{|G_i^*|} \text{ENG}_i
\end{equation}

This fine-tuning process operates on the fundamental criteria of energy levels and coverage ratio, ensuring a delicate balance between sustained energy efficiency and optimal coverage. Notably, its time complexity is \(O(n \cdot m \log n \cdot m)\) , highlighting its efficiency during deployment.


\begin{algorithm}
    \SetAlgoLined
    \SetKwInOut{Input}{Input}
    \SetKwInOut{Output}{Output}
    \SetKw{KwDownTo}{down to}
    
    \Input{Optimal UAV fog node positions, Covered users, Fog node energy levels}
    \Output{Updated fog node positions to maximize the network lifespan}
    
    \BlankLine
    \textcolor{blue}{\textbf{Initialization:}}
    
    Identify crowded areas using UAV fog node positions and determine the covered mobile users within\;
    Rank fog nodes by energy levels (highest to lowest)\;
    Rank fog nodes by coverage ratio (highest to lowest)\;
    
    \BlankLine
    \textcolor{blue}{\textbf{Repositioning Phase:}}
    
    \For{each fog node $F_i$ in sorted high-energy fog nodes}{
        \If{$F_i$ covers few users}{
            Find a neighboring fog node $F_j$ covering high-density users\;
            $F_i$ flies towards the high-density user area covered by $F_j$\;
        }
    }
    
    \For{each fog node $F_k$ in sorted low-energy fog nodes}{
        \If{$F_k$ covers high number of users}{
            Find a neighboring fog node $F_l$ covering low-density users\;
            $F_k$ flies towards the low-density user area covered by $F_l$\;
        }
    }
    
    \textcolor{blue}{\textbf{Output:}}
    
    Updated fog node positions based on optimization for extended network lifespan\;
    
    \caption{UAV Fog Node Repositioning for Maximizing Network Lifespan}
    \label{algo:ECNSA}
\end{algorithm}

\section{Experimental Results and Discussion}
\label{sec.5}

In this section, we delve into the practical implementation of the proposed adaptive WOA algorithm and carry out a set of simulations to assess its performance within the UAV-based fog presented infrastructure we have presented. We start by presenting an overview of the simulation environment. Next, we carefully analyze the results obtained from simulations using the adaptive WOA (Whale Optimization Algorithm) under diverse configurations, comparing these outcomes with PSO (Particle Swarm Optimization), HHO (Harris Hawks Optimization), SCA (Sine Cosine Algorithm), and SMA (Social Memory Algorithm) methods.

\begin{figure}[h]
    \centering
      \includegraphics[width=\linewidth]{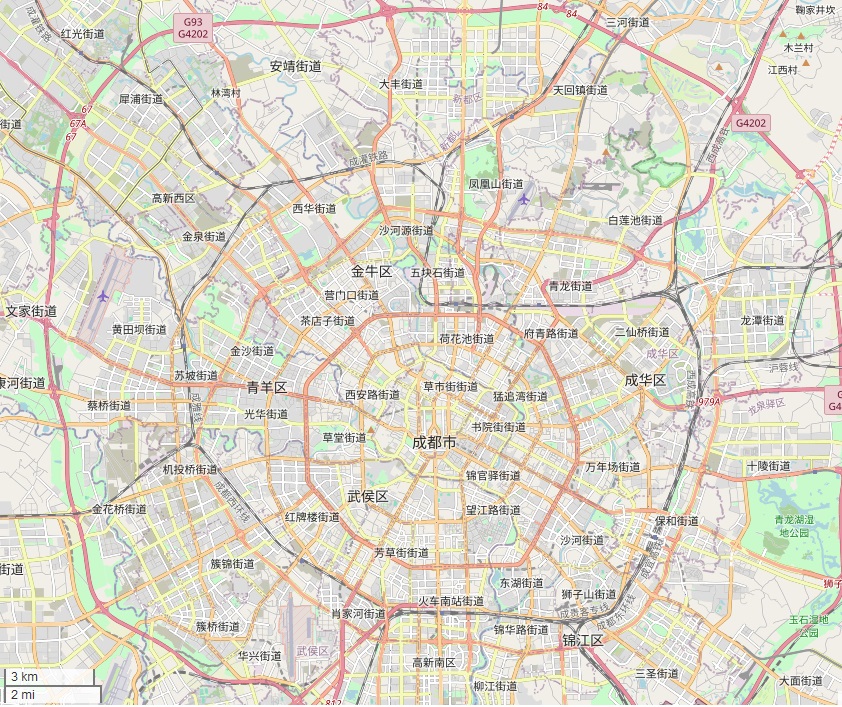}

    \caption{1000m $\times$ 1000m Area Extracted from Sichuan City Using OpenStreetMap}
    \label{fig:area}
\end{figure}

In order to test the efficacy of our algorithmic enhancements, we have opted to develop our solution using the powerful computational capabilities of Matlab 2020Ra. Our experiment is carefully structured to operate within a specific area, precisely a 1000m $\times$ 1000m rectangular zone extracted from Sichua city through OpenStreetMap data figure~\ref{fig:area}. This planned choice allows us to closely replicate real-world scenarios, elevating the practical relevance of our optimization methods. By focusing our simulation within this defined area, our goal is to thoroughly assess the algorithms' performance. We have closely examined different metrics within the spatial context of this specific zone such as convergence, UAV fog nodes connectivity, mobile users coverage, and network lifespan. With utilizing a normal/uniform distribution approach for deploying UAV fog nodes and mobile user nodes. The computations were performed on an AMD Ryzen 75700U processor boasting 8 cores operating at 4.3 GHz, complemented by 8GB of RAM, all within a Windows 11 environment. Our experimental setup affords us the flexibility to replicate the experiment under a range of conditions and constraints, thereby allowing us to carry out the experiment in a controlled environment with predefined settings. Table~\ref{tab:config} presents a comprehensive list of parameters and their respective values, which were determined and refined through a series of initial experiments.

\begin{table}[h]
    \centering

    \begin{tabular}{|c|c|c|}
    \hline
    Parameter & Range & Initial Setting \\ \hline
    UAV Fog Nodes Density & [10, 120] & 45 \\ \hline
    Mobile Users Density & [30, 200] & 120 \\ \hline
    Communication Range & [90, 200] & 100m \\ \hline
    Area Width & 1km & 1km \\ \hline
    Area Height & 1km & 1km \\ \hline
    Altitude & 300-600m & 400m \\ \hline
    Timeframe & 20-60min & 30min \\ \hline
    \end{tabular}
    \caption{Network Configuration Details}
    \label{tab:config}
\end{table}

\subsection{Adaptive WOA EVALUATION with the dynamic environment of UFCS}
This subsection considers the dynamic environment of UFCS. Consider executing 1000 iterations of the Adaptive WOA, HHO, SCA, and SMA algorithms, in which the network topology has a dynamic change each frame $\tau$. We discussed the convergence of these algorithms within timeframe $\tau$, proofing how well the Adaptive WOA and other algorithms converge toward optimal solutions as the number of iterations increases. Convergence analysis can shed light on the speed and stability of convergence and provide insights into the algorithms' efficiency in reaching near-optimal or optimal solutions within a given problem.

\subsubsection{Evaluating Convergence: A Comparative Analysis of Applied Algorithms}
\begin{figure*}[h]
    \centering
    \begin{subfigure}[b]{0.48\linewidth}
        \centering
        \includegraphics[width=\linewidth]{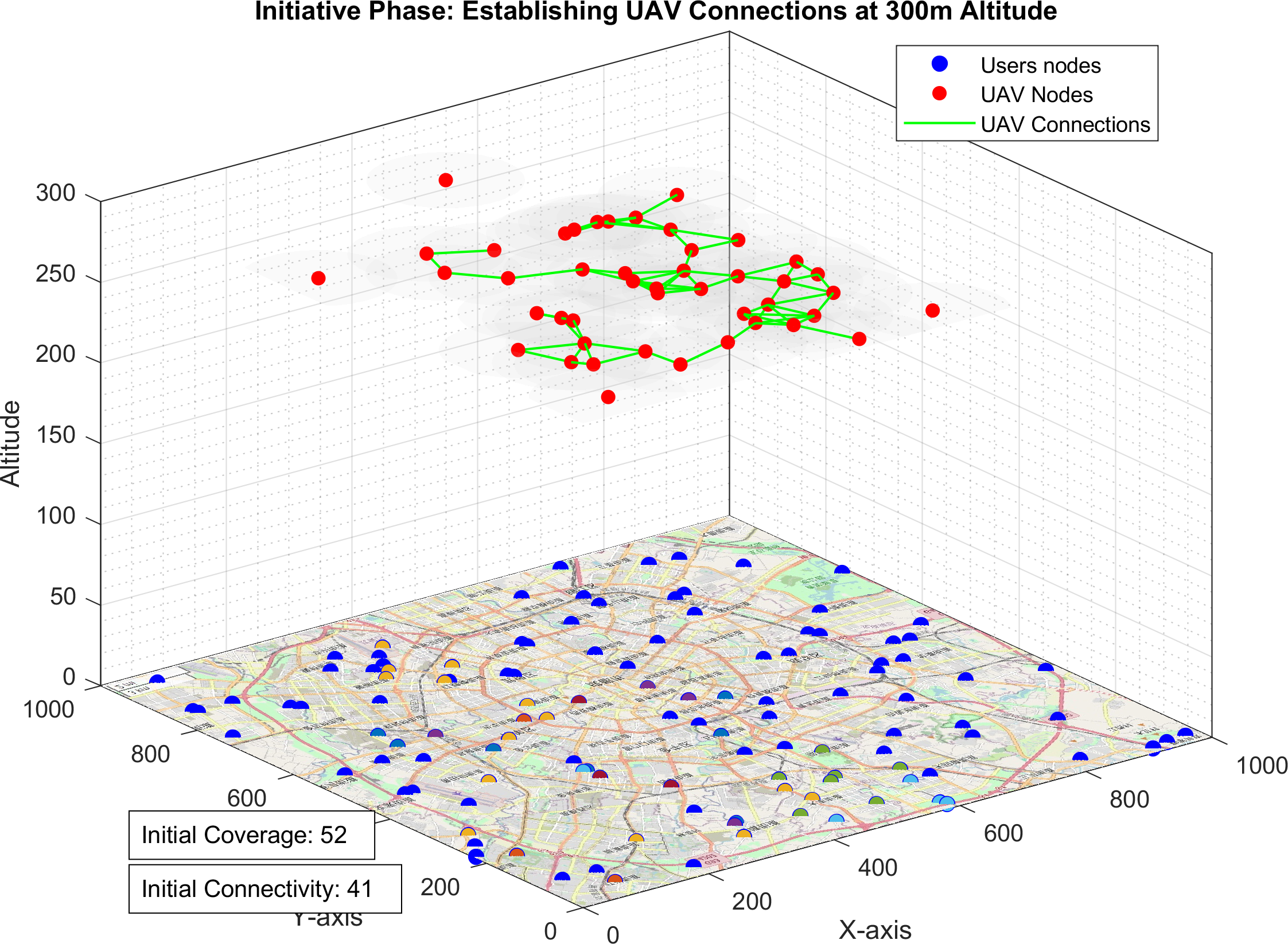}
        \caption{Initial Deployment - 41 Connected UAV-Fog Nodes Covering 52 Users}
        \label{fig:two_instance_subfig1}
    \end{subfigure}
    \quad
    \begin{subfigure}[b]{0.48\linewidth}
        \centering
        \includegraphics[width=\linewidth]{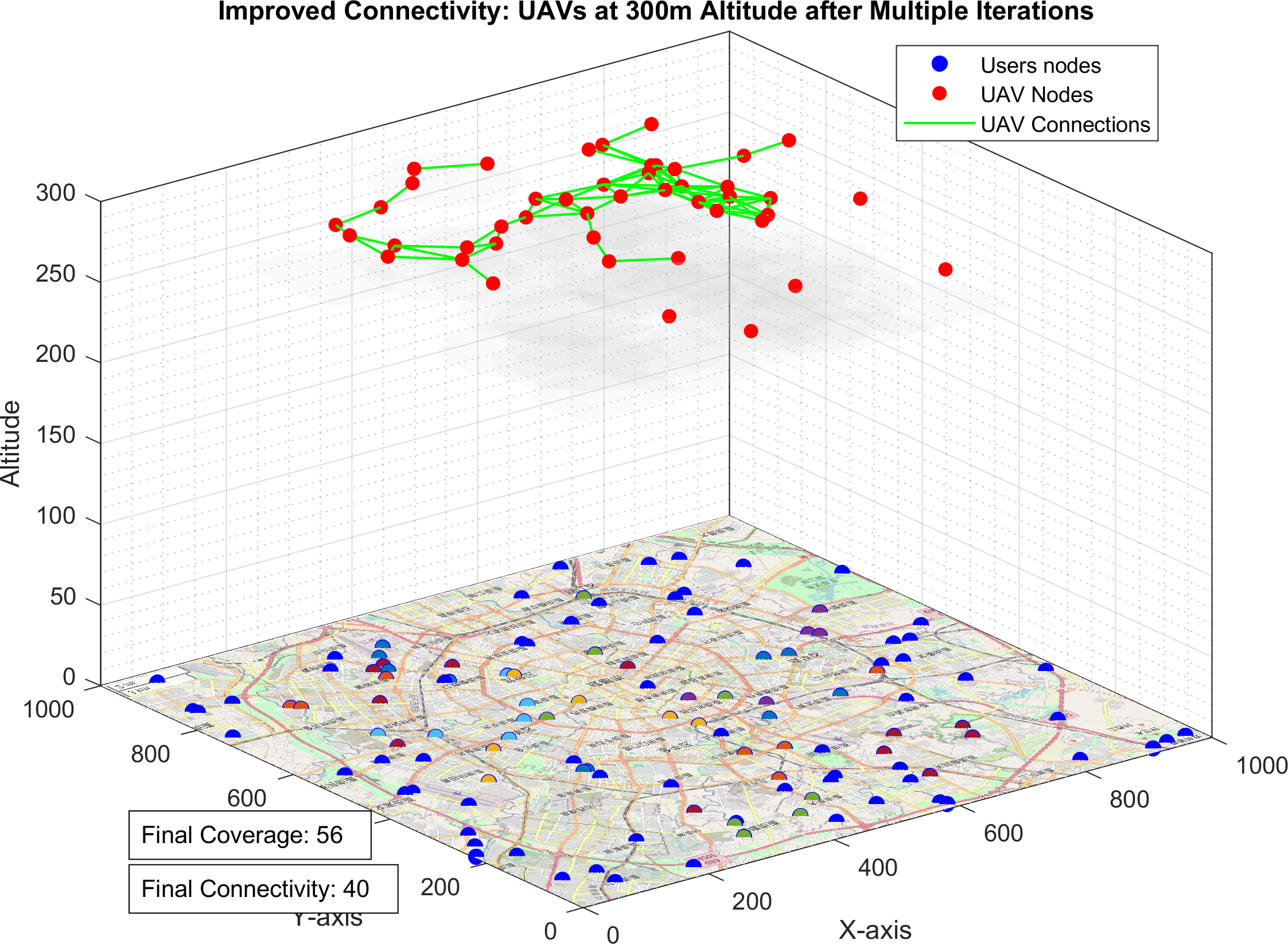}
        \caption{Improved Deployment - 40 UAV-Fog Nodes Covering 56 Users}
        \label{fig:two_instance_subfig2}
    \end{subfigure}
    \caption{Comparison of Deployments}
    \label{fig:two_instance_fig}
\end{figure*}

Figure~\ref{fig:two_instance_fig} presents a visual instance of the problem we're investigating. In this scenario, we have two types of nodes: UAV-fog nodes, shown as solid red circles, and nodes for affected mobile users, shown as solid blue circles. These nodes are randomly placed in a 1000x1000 unit square. If two fog nodes are within communication range, they are connected by a green link, as seen between fog nodes in the figure. When a user node falls within the communication range of a fog node, it's represented in a distinct color from the blue nodes, which signify uncovered clients. Figure~\ref{fig:two_instance_subfig1} displays 52 covered users with 41 connected UAV fog nodes, while Figure ~\ref{fig:two_instance_subfig2} shows an improvement after multiple iterations, covering 56 users with 40 UAV fog nodes. This improvement signifies that adopting the Adaptive WOA algorithm enhances both coverage and connectivity. Importantly, using this algorithm doesn't just improve performance but also reduces the deployment costs. By employing fewer UAV fog nodes, we can cover more mobile users efficiently with low cost.

\begin{figure}[h]
    \centering
    \includegraphics[width=1.0\linewidth]{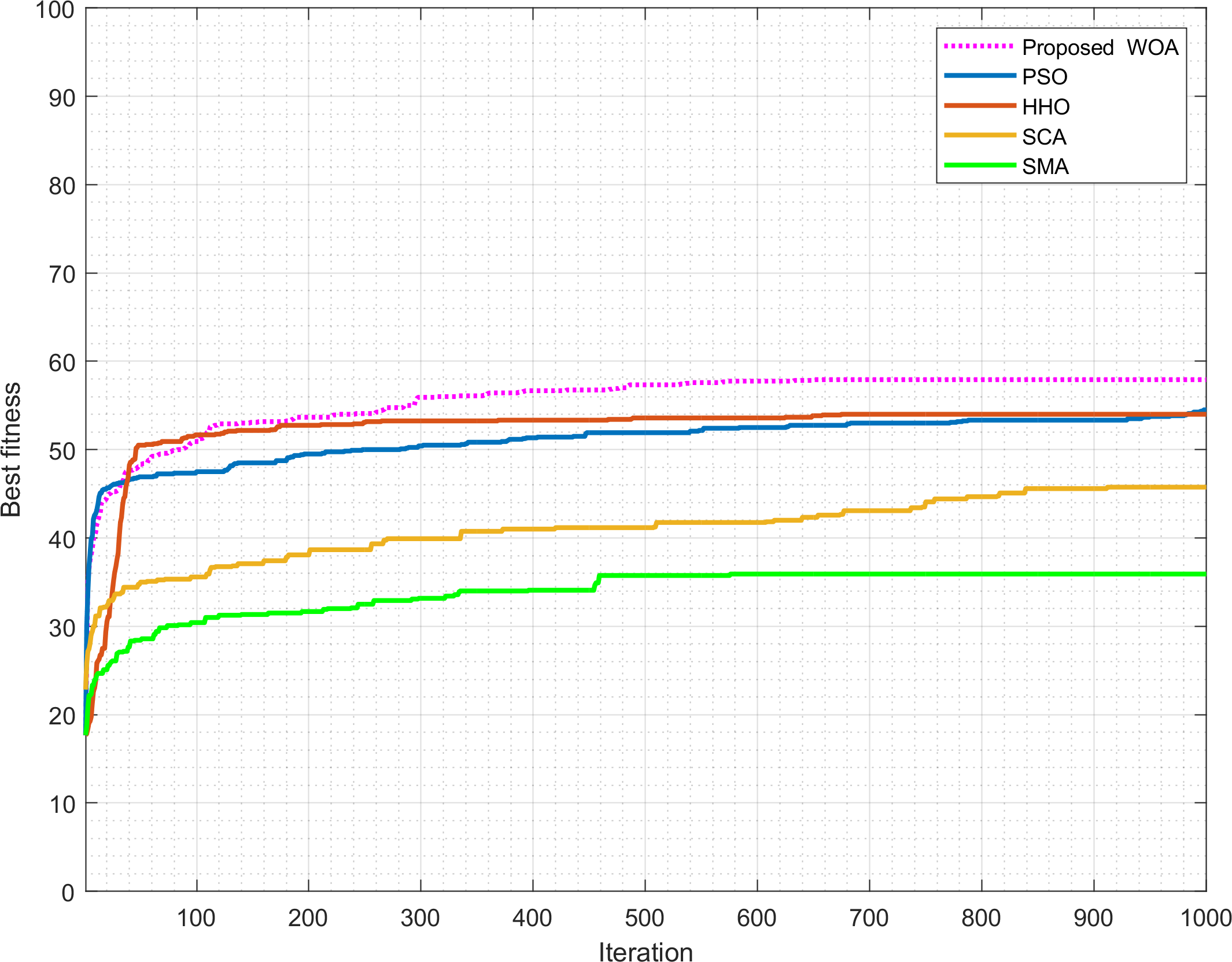}
    \caption{Convergence Analysis of Optimization Algorithms}
    \label{fig:convergence}
\end{figure}

The convergence analysis is illustrated in Figure~\ref{fig:convergence}, where the x-axis represents the iteration count, and the y-axis signifies the best fitness value at each iteration. The dashed and solid lines correspond to the outcomes of the HHO, SCA, SMA, and the proposed Adaptive WOA, respectively. Initially, all algorithms exhibited significant changes during the experiment, but these changes gradually reduced as the iterations progressed. In contrast, the HHO, SCA, SMA algorithms face challenges related to premature convergence, increasing the likelihood of getting trapped in a local optimum rather than a global one. An algorithm's overall performance is significantly influenced by its rate of progress. As depicted in Figure~\ref{fig:convergence}, the proposed Adaptive WOA algorithm demonstrates excellent potential and achieves faster convergence compared to the other algorithms after a certain number of iterations, ultimately reaching an optimal fitness function value.

\begin{figure*}[h]
    \centering
    \begin{subfigure}[b]{0.4\linewidth}
        \centering
        \includegraphics[width=\linewidth]{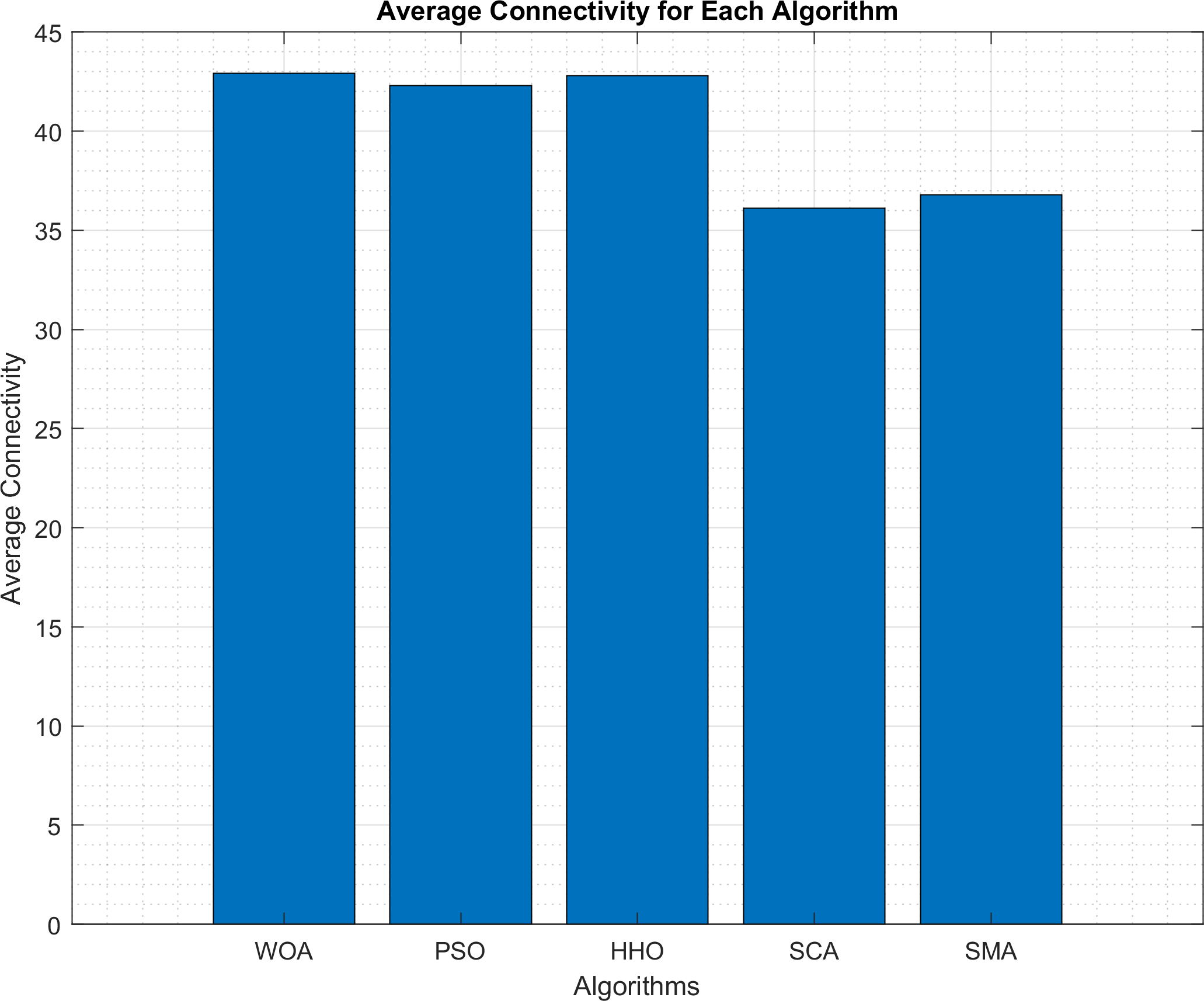}
        \caption{Network Connectivity Analysis: Adaptive WOA, PSO, HHO, SCA, SMA}
        \label{fig:network_connectivity}
    \end{subfigure}
    \quad
    \begin{subfigure}[b]{0.4\linewidth}
        \centering
        \includegraphics[width=\linewidth]{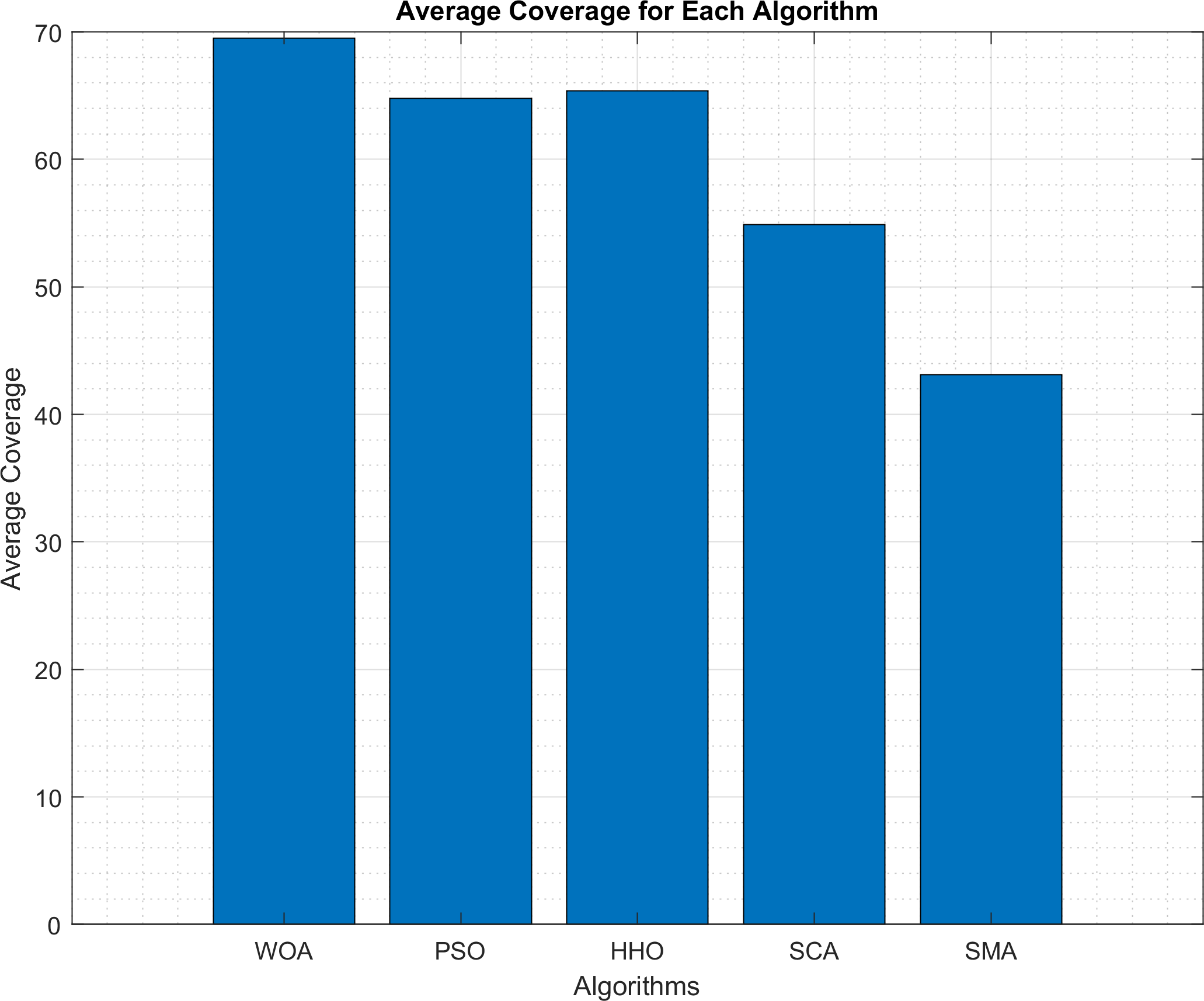}
        \caption{Network Coverage Analysis: Adaptive WOA, PSO, HHO, SCA, SMA}
        \label{fig:network_coverage}
    \end{subfigure}
    \caption{UAV Fog nodes Connectivity and Mobile users Coverage Comparison}
    \label{fig:connectivity_coverage}
\end{figure*}

Furthermore, for a more comprehensive understanding of the refinements made to the optimization objectives, we conducted a performance evaluation comparing the proposed Adaptive WOA optimization algorithm and other algorithms with different key parameters. First, this evaluation was based on the following assessment criteria: UAV nodes connectivity (NC) and mobile user coverage (NCV) as shown in Figure~\ref{fig:connectivity_coverage}. We assess the performance of our proposed Adaptive WOA (Adaptive Whale Optimization Algorithm) in comparison to well-established algorithms such as PSO (Particle Swarm Optimization), HHO (Harris Hawks Optimization), SCA (Sine Cosine Algorithm), and SMA (Social Memory Algorithm) for the defined objective function H (equation \ref{eq:H_X}). In this part, our evaluation shows the UAV fog nodes connectivity and mobile user coverage finding with different scenarios. Figure \ref{fig:connectivity_coverage} visually represents our algorithm's ability to achieve superior coverage of affected mobile users while utilizing fewer UAV fog nodes. Notably, in Figure~\ref{fig:network_connectivity} and Figure~\ref{fig:network_coverage}, our algorithm demonstrates a remarkable 43\% connectivity rate, encompassing 70\% of the target population. Further analysis reveals that while the PSO and HHO algorithms achieve comparable connectivity rates to our proposed Adaptive WOA, they cover fewer mobile users. Specifically, the PSO and HHO achieve coverage of 65\% and 64\% of mobile users, respectively, with 43\% UAV-connected nodes. Contrastingly, our Adaptive WOA achieves an impressive 70\% coverage with the same 43\% UAV connectivity.

\subsection{Assessing the Efficiency of Our Proposed Fitness Function}

This subsection revolves around evaluating the effectiveness of the proposed functions, particularly emphasizing the role of the fitness function $H$, where this function plays significant importance in assessing various metrics, including user coverage and UAV fog nodes connectivity, within the scope of the WHO algorithm. We employ identical parameters across multiple experiments, to determine how efficiently the Adaptive WOA algorithm achieves the optimal value as defined by $H$. Furthermore, subsequent to this initial phase, we delve into an analysis of an energy-conscious rearrangement strategy for UAV fog nodes. This analysis expands beyond the evaluation of user’s coverage and network connectivity, encompassing an assessment of the network's durability and overall efficiency.

\begin{figure*}[h]
    \centering
    \begin{subfigure}{0.45\textwidth}
        \centering
        \includegraphics[width=\linewidth]{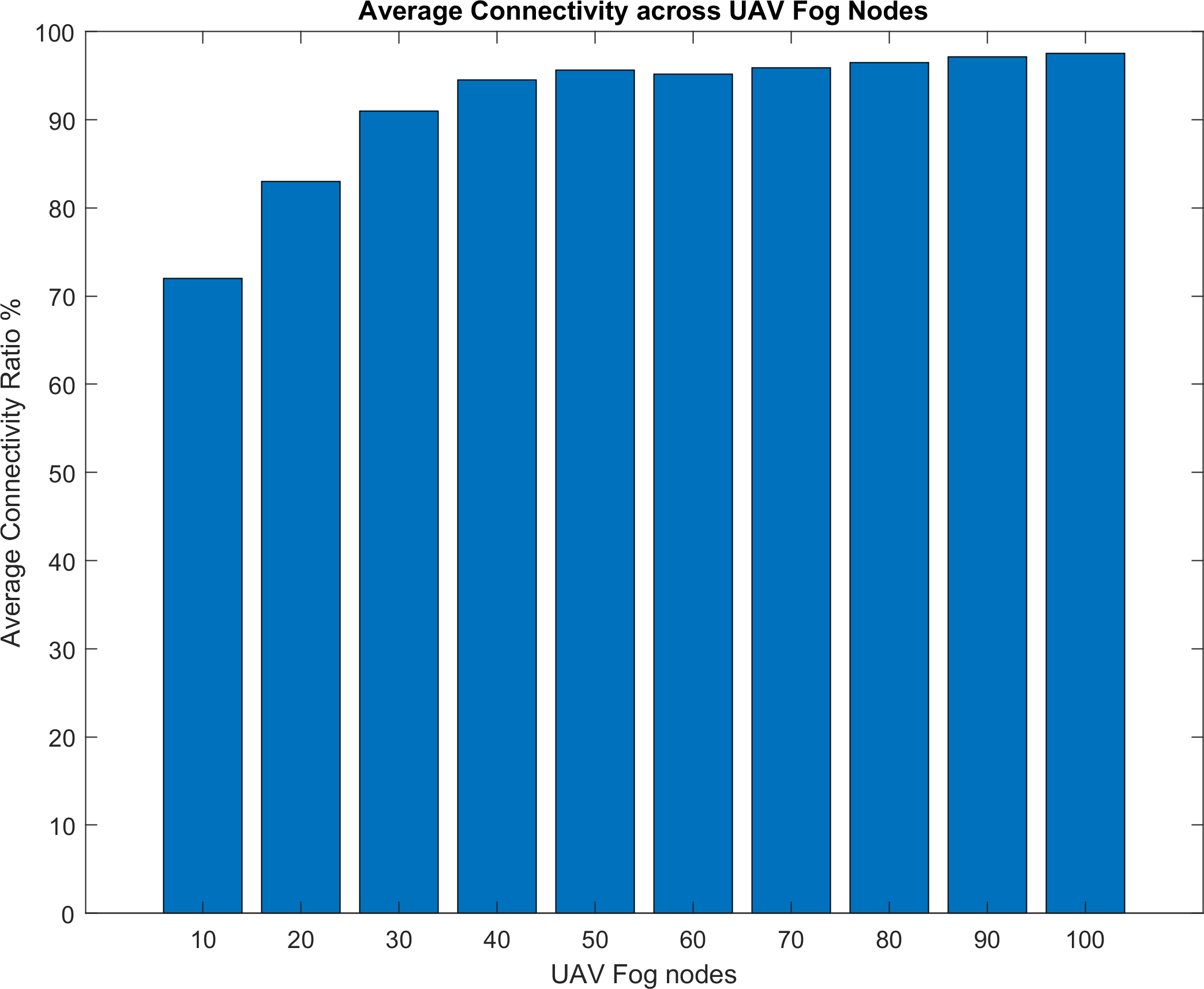}
        \caption{Connectivity}
        \label{fig:sub1}
    \end{subfigure}
    \hfill
    \begin{subfigure}{0.45\textwidth}
        \centering
        \includegraphics[width=\linewidth]{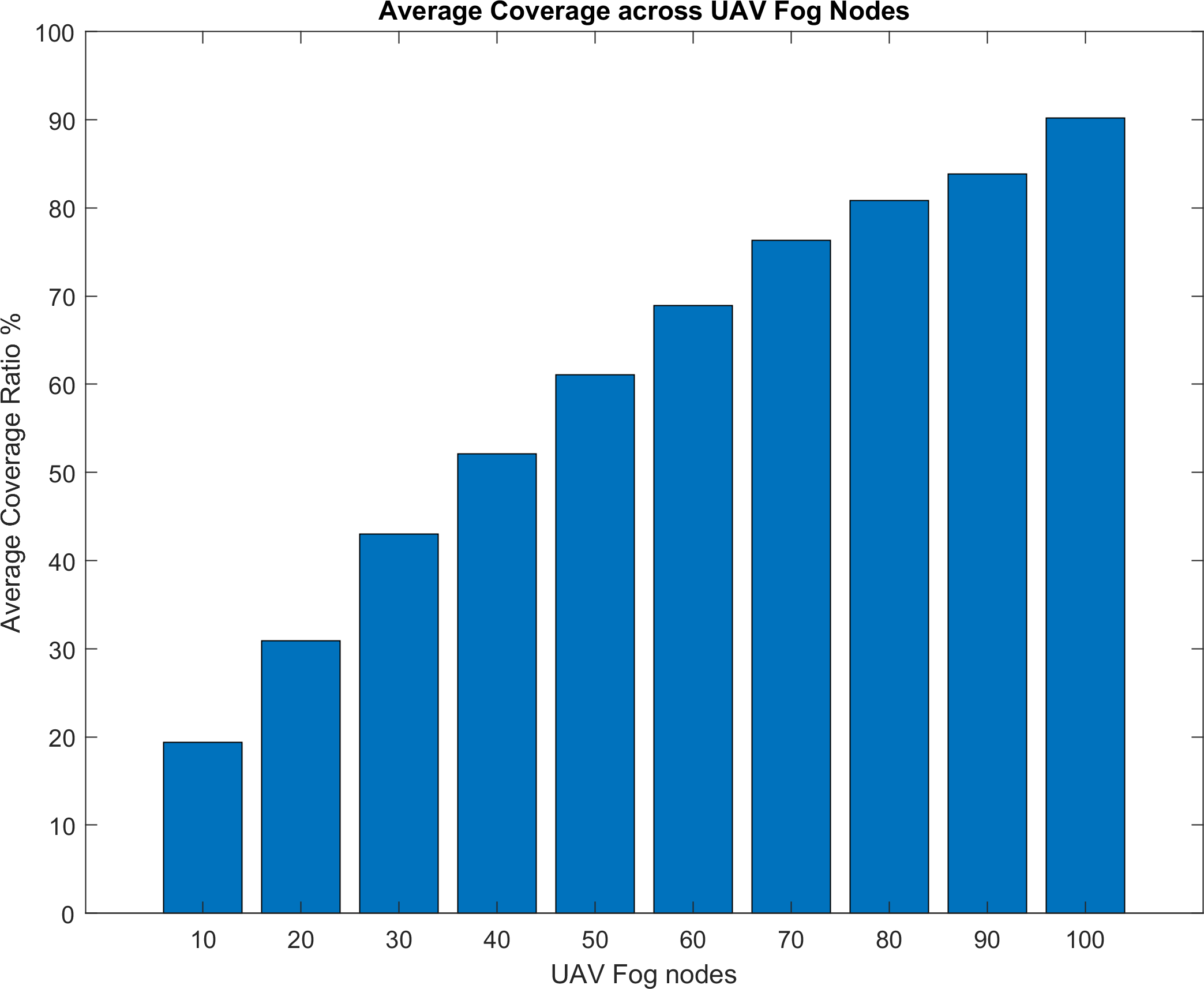}
        \caption{Coverage}
        \label{fig:sub2}
    \end{subfigure}
    \caption{Effect of Increased UAV Fog Nodes on Network Coverage and Connectivity}
    \label{fig:uav_nodes_increase}
\end{figure*}

In Figure \ref{fig:uav_nodes_increase}, the escalation of UAV fog nodes from 10 to 100 corresponds to a simultaneous increase in both the percentage of covered users and the connectivity among these nodes, amplifying the network's overall efficacy. Additionally, the network topology maintains consistent connectivity, highlighting its robust stability. The inclusion of extra nodes facilitates a more precise allocation of resources, addressing areas or users that previously encountered weaker coverage. Notably, the expanded count of UAV fog nodes substantially widens the network's coverage area, directly amplifying connectivity and accessibility to services for users across the network.

\begin{figure*}[h]
    \centering
    \begin{subfigure}{0.45\textwidth}
        \centering
        \includegraphics[width=\linewidth]{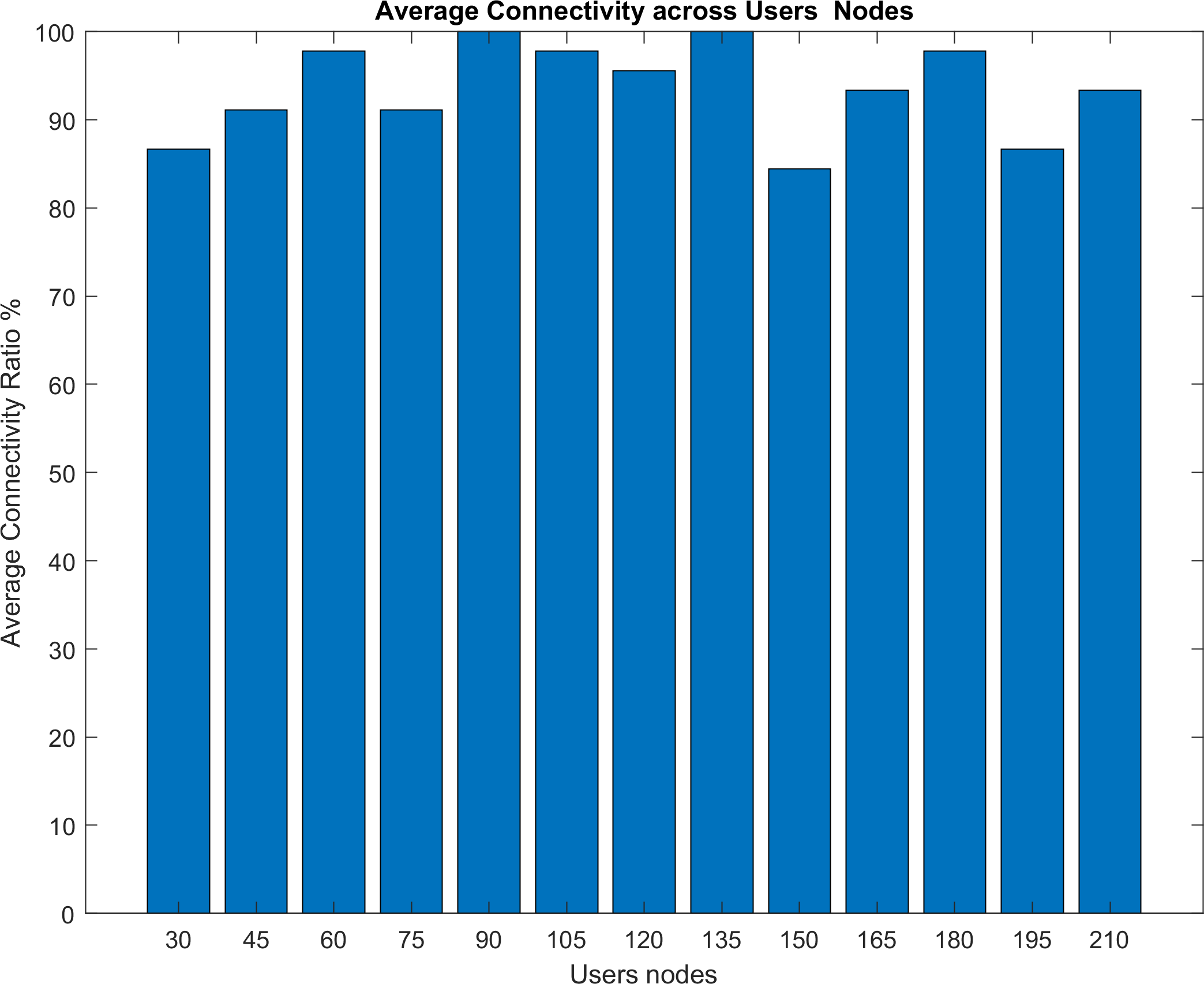}
        \caption{Connectivity}
        \label{fig:user_sub1}
    \end{subfigure}
    \hfill
    \begin{subfigure}{0.45\textwidth}
        \centering
        \includegraphics[width=\linewidth]{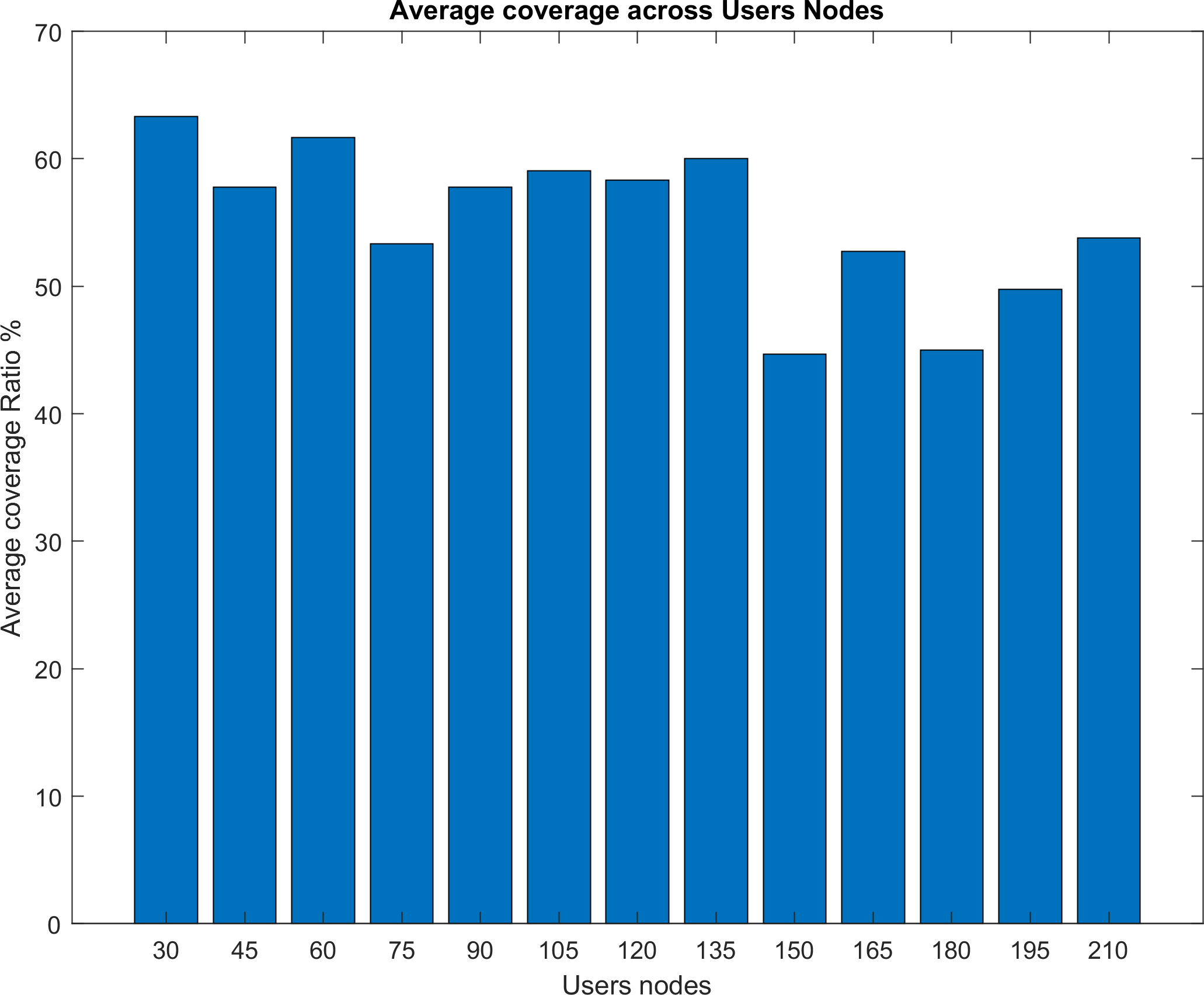}
        \caption{Coverage}
        \label{fig:user_sub2}
    \end{subfigure}
    \caption{Effect of Increased UAV Fog Nodes on Network Coverage and Connectivity}
    \label{fig:user_nodes_increase}
\end{figure*}

The findings presented in Figure \ref{fig:user_nodes_increase} explore how changes in the number of mobile users, ranging from 50 to 275, impact the connectivity among UAV fog nodes and the coverage offered to mobile users within the network. In Figure \ref{fig:user_sub1}, where the focus is on the connected UAV fog nodes, the percentage remained fairly consistent, ranging between 70\% and 90\%. This range indicates that the network consistently established connections between available UAV fog nodes to accommodate more mobile users, suggesting efforts to link these nodes together to serve a greater number of mobile users efficiently. In parallel, Figure \ref{fig:user_sub2} underscores the stability in the percentage of users covered by the network, maintaining a range between 70\% and 93\%, despite fluctuations in the mobile user ratio. This stability persists regardless of variations in mobile users’ density due to the uniform distribution over the target area.

\begin{figure*}[h]
    \centering
    \begin{subfigure}{0.45\textwidth}
        \centering
        \includegraphics[width=\linewidth]{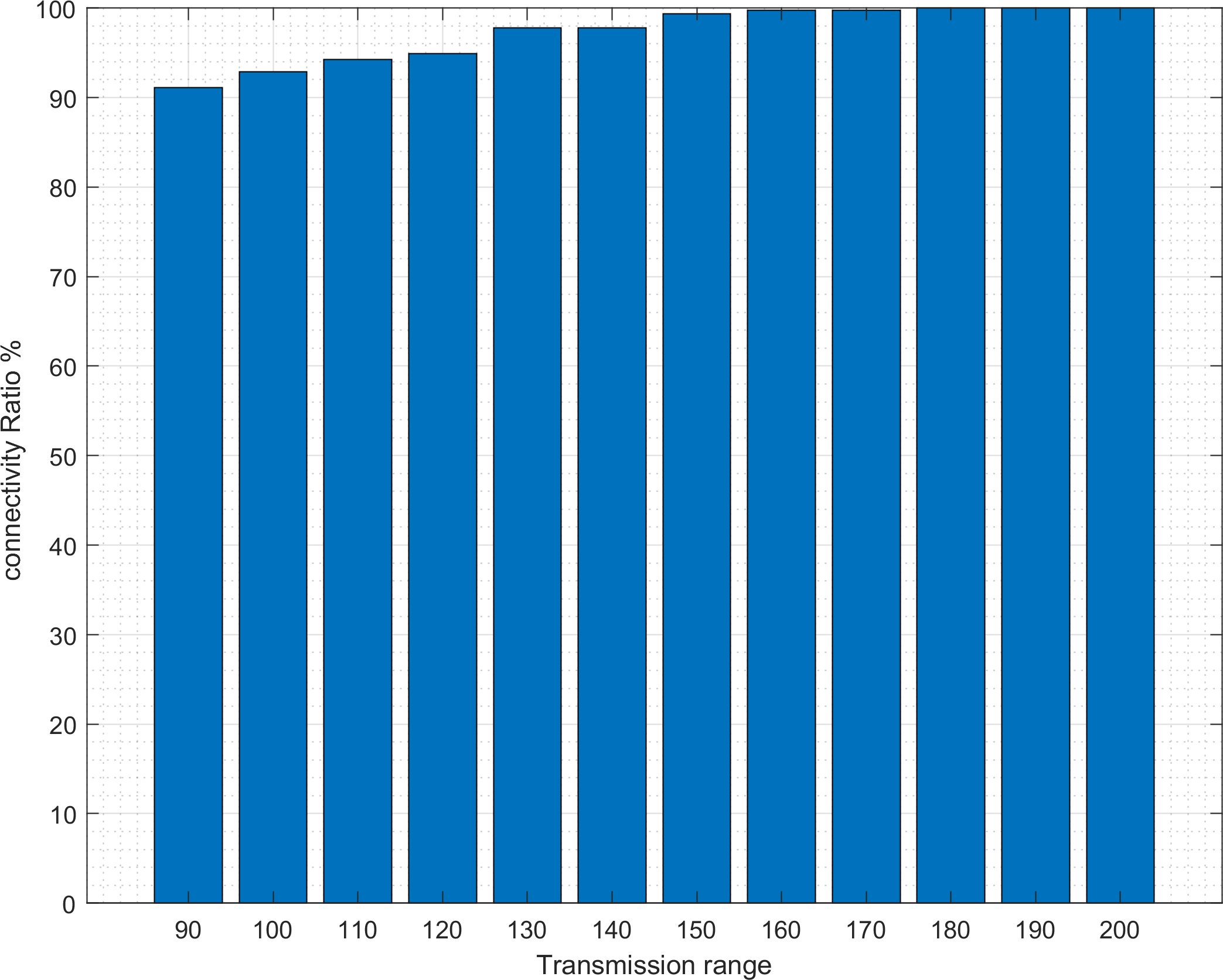}
        \caption{Connectivity}
        \label{fig:utransmission_range_sub1}
    \end{subfigure}
    \hfill
    \begin{subfigure}{0.45\textwidth}
        \centering
        \includegraphics[width=\linewidth]{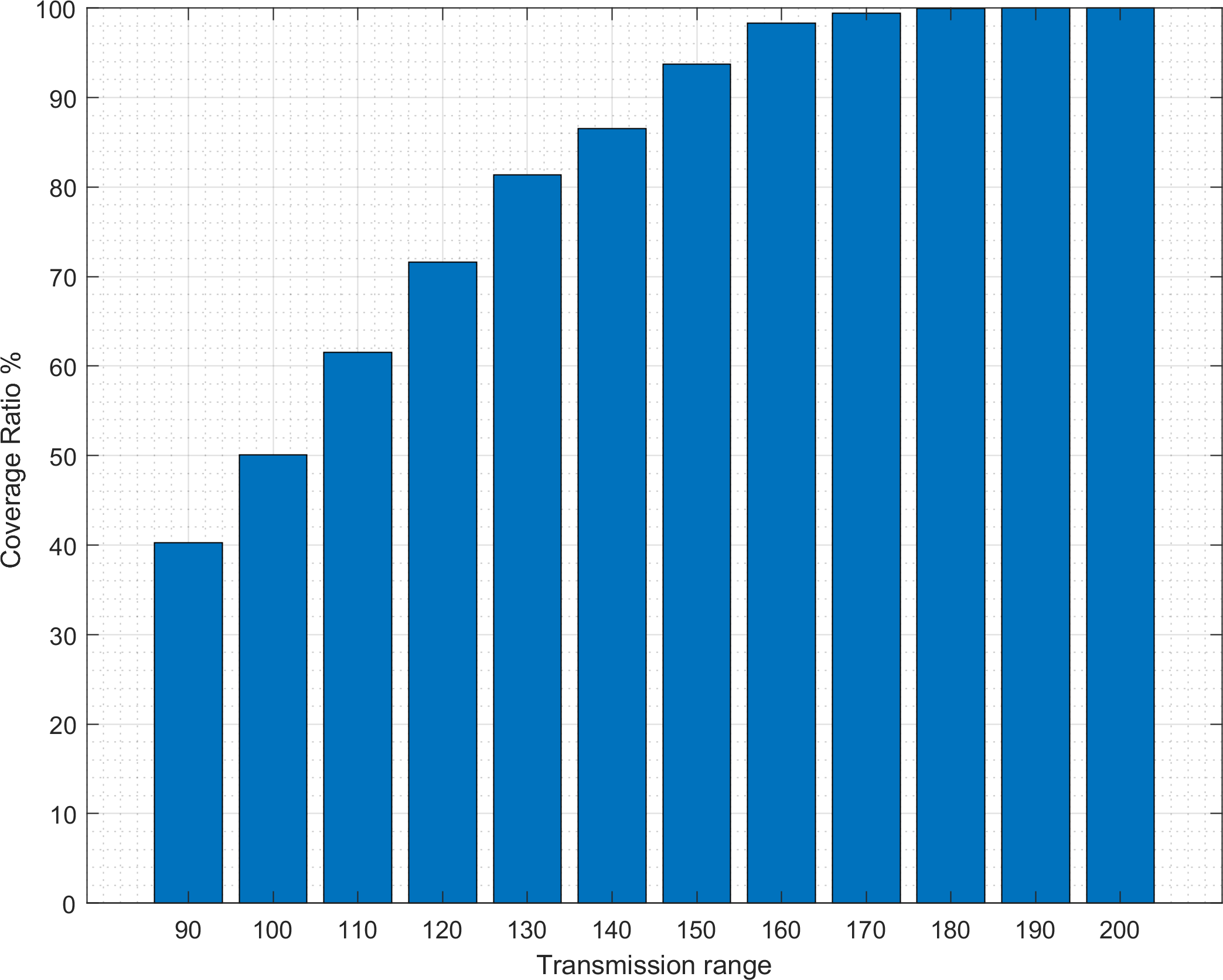}
        \caption{Coverage}
        \label{fig:transmission_range_sub2}
    \end{subfigure}
    \caption{Effect of Increased Transmission Range on Network Coverage and Connectivity}
    \label{fig:transmission_range}
\end{figure*}

Figure \ref{fig:transmission_range} shows the relationship between the transmission range of UAV fog nodes and the coverage of mobile users, and the UAV fog nodes' connectivity. By adjusting the range of the communication from 90 meters up to 200 meters, respectively. The results demonstrate that as the transmission range increases, there's a direct improvement in both the coverage area and the connectivity of the network. For example, Figure \ref{fig:utransmission_range_sub1} shows that when the communication range increases, all fog nodes start attempting to connect with each other, connecting the isolated sub-networks to the largest network, forming a unified network. Hence, expanding the transmission range essentially means that the UAV fog nodes can communicate and reach a larger area. This extension in their reach leads to a broader coverage area, allowing them to serve more mobile users within that expanded range. Consequently, In Figure \ref{fig:transmission_range_sub2}, the illustration depicts the effects of varying communication ranges on the mobile users' coverage ratio. Particularly noticeable is that when the communication range exceeds 150 meters, the majority of mobile nodes within the network come under coverage. This implies that Nearly all of these mobile nodes fall within the communication range established by the network's infrastructure.

\begin{figure*}[h]
    \centering
    \begin{subfigure}[b]{0.45\linewidth}
        \centering
        \includegraphics[width=\linewidth]{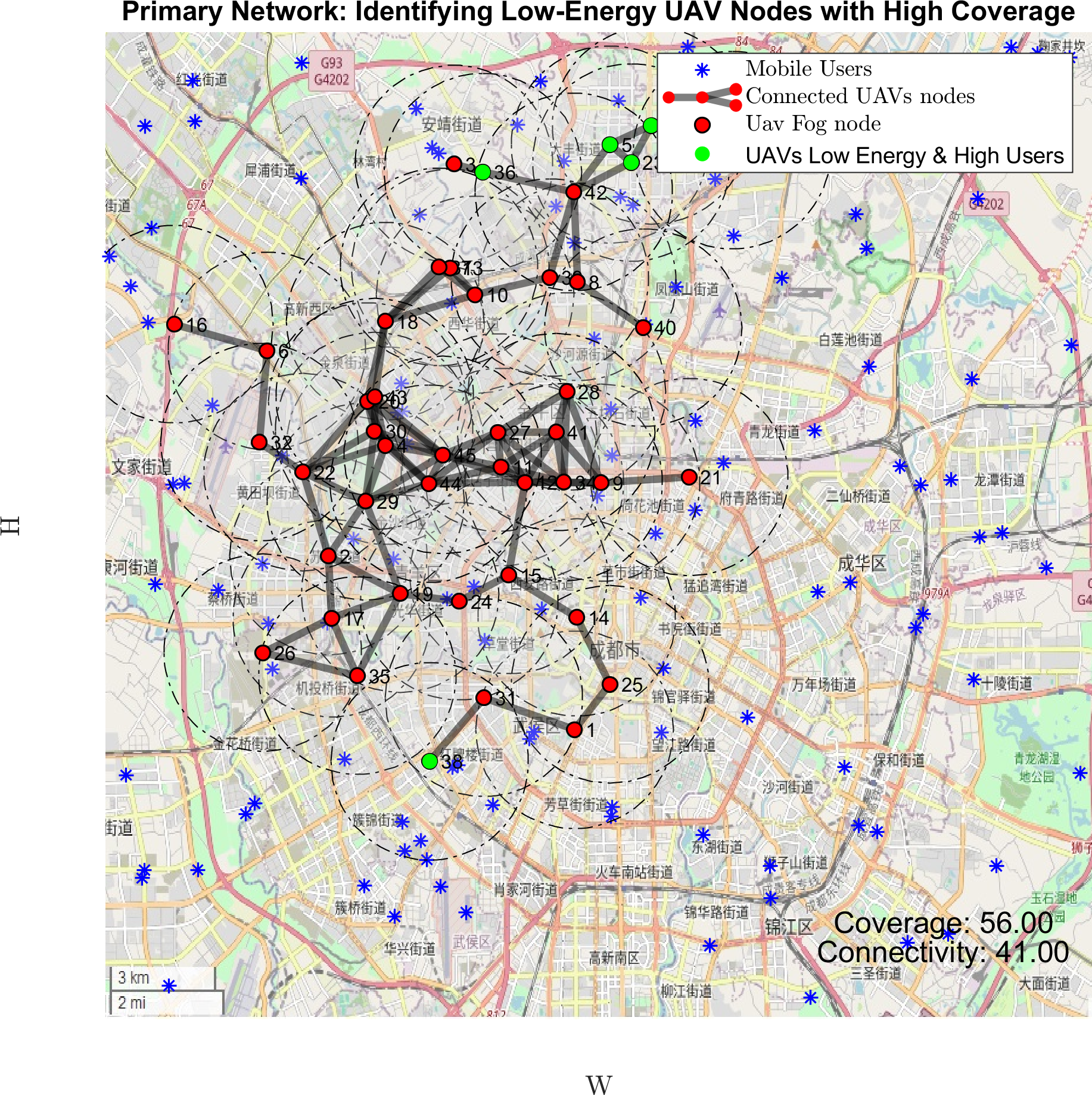}
        \caption{Initial Network State: Optimal Connectivity and Coverage}
        \label{fig:subfig1}
    \end{subfigure}
    \quad
    \begin{subfigure}[b]{0.45\linewidth}
        \centering
        \includegraphics[width=\linewidth]{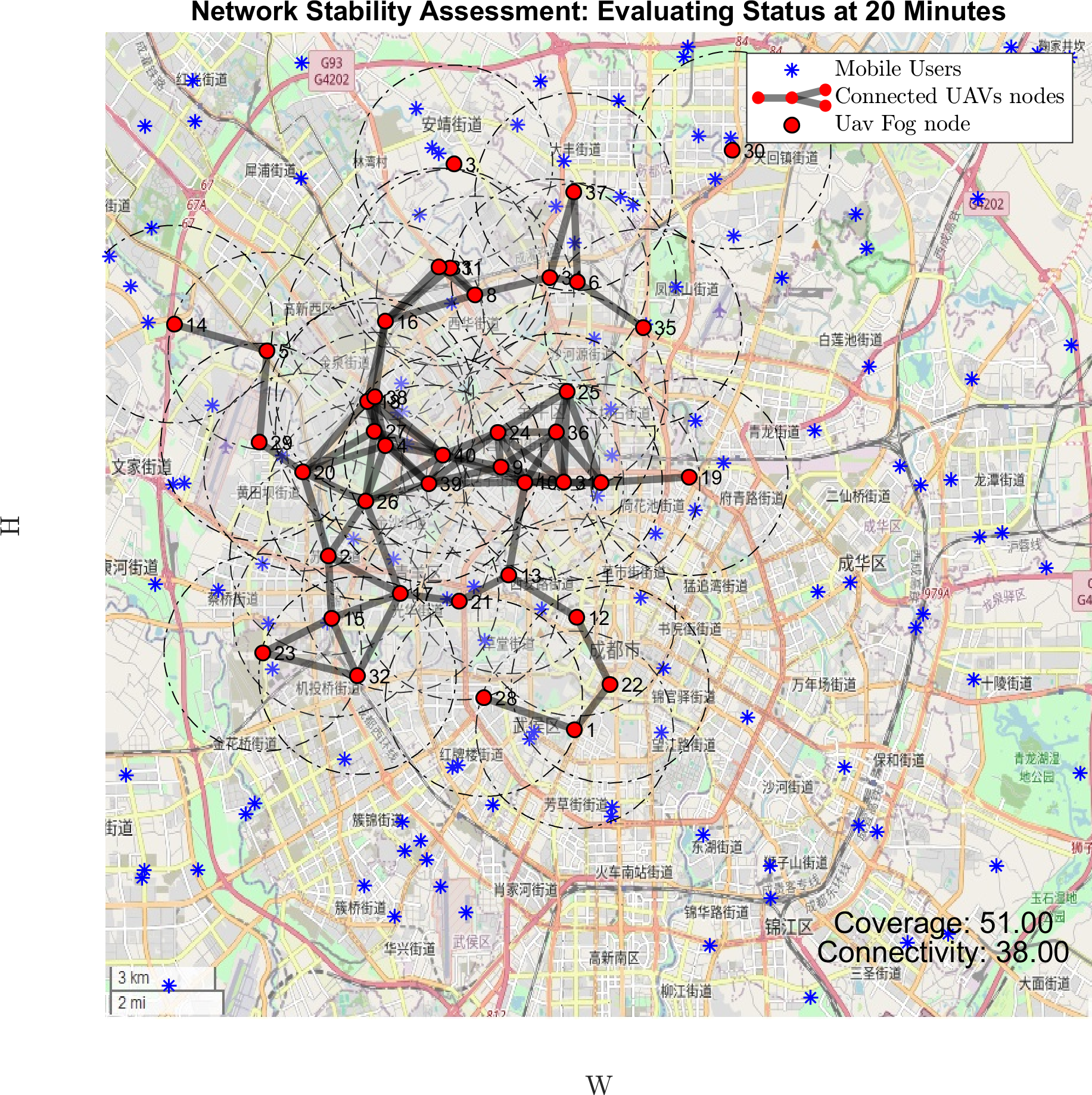}
        \caption{Network Evolution after 20 Minutes: Decline in Connectivity and Coverage Ratios}
        \label{fig:subfig2}
    \end{subfigure}
    \vskip\baselineskip
    \begin{subfigure}[b]{0.45\linewidth}
        \centering
        \includegraphics[width=\linewidth]{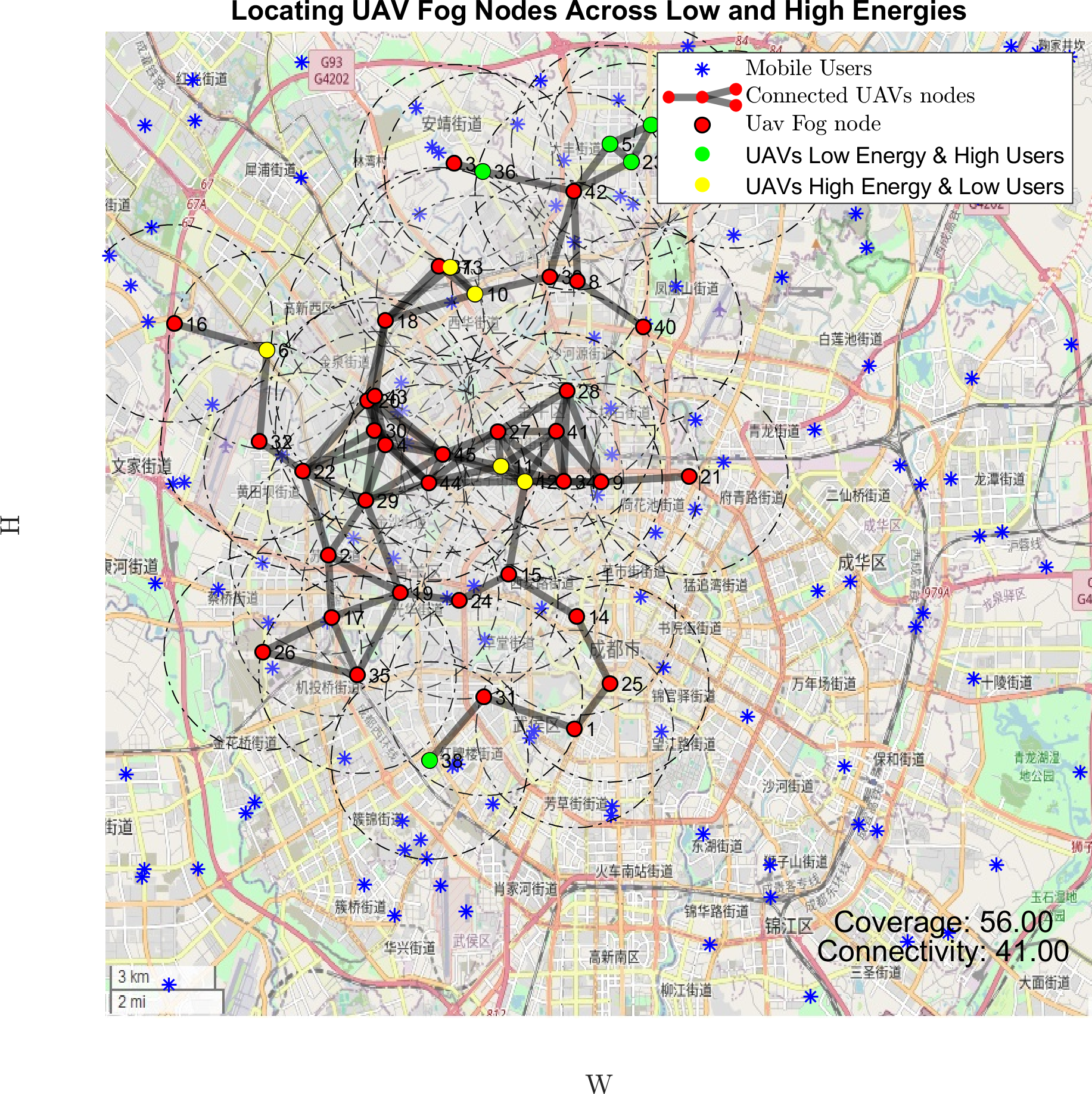}
        \caption{Optimization (ECNSA): Swapping for Enhanced Coverage}
        \label{fig:subfig3}
    \end{subfigure}
    \quad
    \begin{subfigure}[b]{0.45\linewidth}
        \centering
        \includegraphics[width=\linewidth]{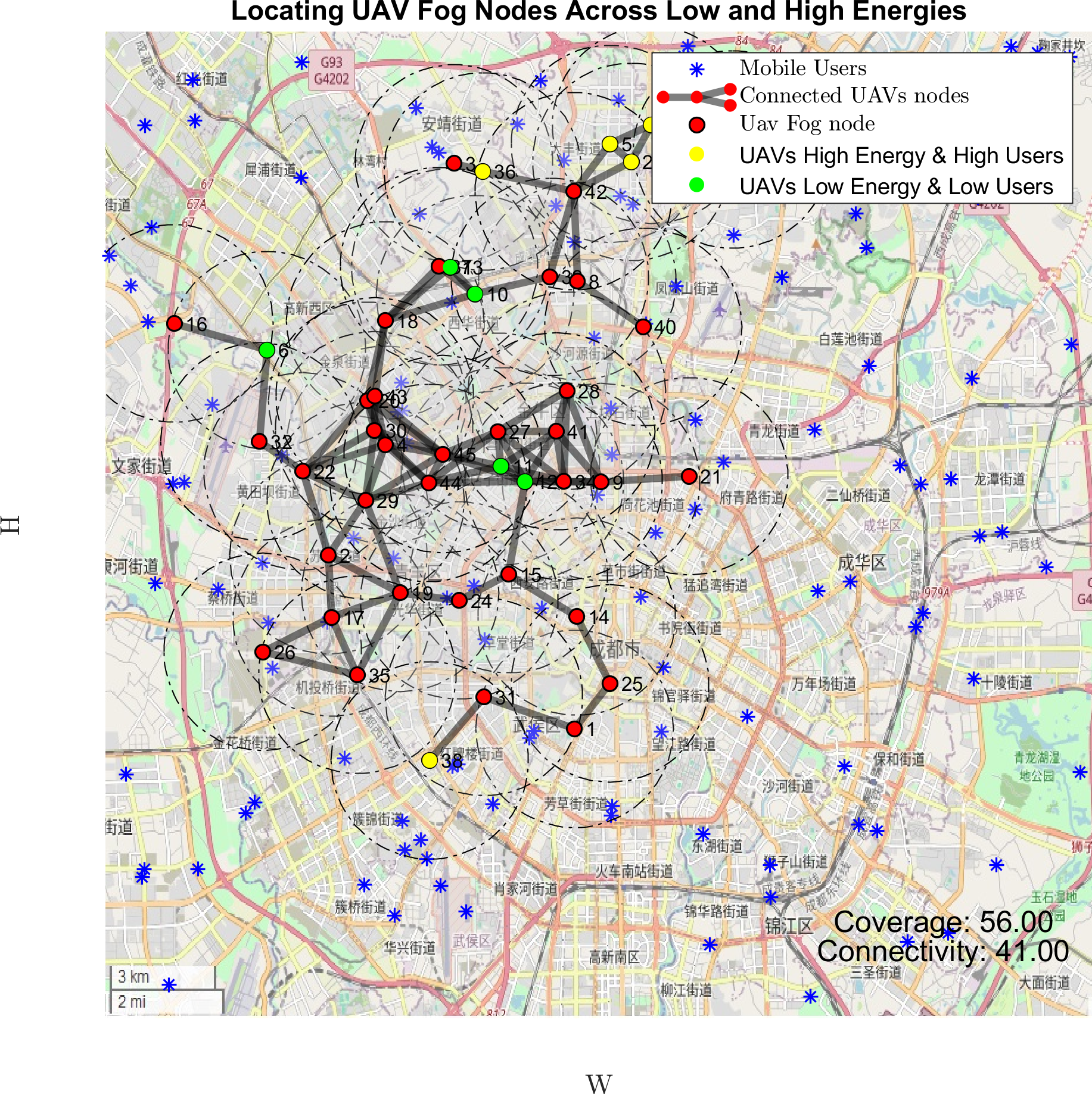}
        \caption{Stable Network: Sustained Connectivity and Coverage}
        \label{fig:subfig4}
    \end{subfigure}
    \caption{Analysis of Network Connectivity and Coverage Over Time}
    \label{fig:network_lifespan}
\end{figure*}

Furthermore, in order to assess the network's lifespan, an evaluation of both coverage and connectivity over time is imperative. This analysis offers a deeper understanding of their evolution and influence on the network's longevity. Hence, Figure \ref{fig:network_lifespan} depicts an analysis of UAV fog nodes connectivity and mobile users coverage, aiming to maximize the network's lifespan by minimizing energy usage. These analyses are derived from the topologies obtained via the WAO algorithm, aligning with our objective functions $H$ (\ref{eq:H_X}). Figures \ref{fig:subfig1},\ref{fig:subfig2},\ref{fig:subfig3} and \ref{fig:subfig4} showcase the network state across different instances. In Figure \ref{fig:subfig1}, the initial network state of optimal determination of connectivity and coverage is illustrated, the highlighted green UAV nodes within represent the nodes that cover a high user ratio while operating on low energy. In Figure \ref{fig:subfig2}, we observed after a duration of 20 minutes, there is a decline in both connectivity and coverage ratios from 56\% and 41\% to 51\% and 38\% respectively, leading to network fragmentation. This drop reflects changes in the network state, particularly influenced by the limited energy of UAV nodes and the user density. Where high users can lead to high traffic \cite{Aung2023,Zhang2018,Aung2020}, consuming high energy of the UAV fog nodes. Our post-optimization algorithm, ECNSA, leverages a swapping approach to strategically reposition UAV nodes. Where nodes with high energy are redirected to areas covered by nodes with lower energy levels. For example, in Figure \ref{fig:subfig3}, the green UAV nodes represent critical areas with high user concentrations compared to others. The remaining fog nodes are identified as potential candidates for initiating the swapping process, while the yellow UAV fog nodes mark optimal positions for the optimization process. Notably, in Figure \ref{fig:subfig4}, the network demonstrates stability for over an hour, with sustained connectivity and coverage. Furthermore, while the UAV consuming their energy to provide covering service to users the optimization process, we always keep periodically locating for the best affectation to extend the network lifespan.

\section{Conclusion}
\label{sec.6}

In this study, we introduced a cutting-edge three-tier dynamic UAV-based fog integration infrastructure, focusing on the deployment process of UAV fog nodes to deliver critical services in disaster response scenarios. And, we extracted a 1 km rectangular zone from Sichuan city, replicating real-world scenarios and elevating the relevance of our optimization methods. This deliberate choice enabled us to closely simulate disaster response scenarios, emphasizing the practical applicability and relevance of our research outcomes. Our investigation centered on the intricate dynamics of transmission, coverage, and wireless communication within this framework. To address the challenges, we devised a dual-focused solution; the Adaptive WOA Meta-heuristic algorithm, designed to optimize UAV fog node deployment by maximizing connectivity and coverage, and the ECNSA algorithm, precisely tuned to factor in energy considerations and population density. The Collaboration between these algorithms significantly enhanced network throughput while simplifying the original problem into two distinct subproblems. The first subproblem focused on achieving optimal network connectivity and coverage, while the second utilized node and environmental coverage information to iteratively position UAV networks, effectively extending the network's lifespan.  Our extensive simulations yielded compelling results, showcasing the advantage of our proposed experimental scheme over existing methods. Notably, the Adaptive WOA algorithm exhibited rapid convergence and low computational complexity, outperforming state-of-the-art baselines. Additionally, the ECNSA algorithm proved key in extending the network's operational duration. Moreover, we formulated a comprehensive mathematical model defining connectivity and coverage conditions, accompanied by an objective function tailored to optimize UAV fog nodes. This model focuses maximizing connectivity, coverage, and extend network life duration. In future work, we will investigate the integration of predictive analytics and machine learning algorithms to improve decision-making and resource allocation. These technologies could optimize the network's response by predicting disaster patterns, resource demands, or optimal node placements based on historical data or real-time inputs.

\bibliographystyle{IEEEtran}
\bibliography{sample}

\end{document}